\documentclass[a4paper, 11pt,  numbers=noenddot, bibliography=totoc, abstracton]{scrartcl}

\usepackage{setspace}
\usepackage[ngerman,english]{babel}
\usepackage[T1]{fontenc}
\usepackage[utf8]{inputenc}
\usepackage{tabto}
\usepackage[margin=1in]{geometry}
\usepackage{graphicx} 
\usepackage{floatflt} 
\usepackage{amsmath}  
\usepackage{amssymb}  
\usepackage{amsthm}   
\usepackage{paralist} 
\usepackage{color}	
\usepackage{scrlayer-scrpage}
\usepackage{bbm} 
\usepackage{tikz} 
\usepackage{tikz-cd} 
\usepackage{array}
\usepackage{pgfplots} 
\usepackage[section]{placeins} 
\usepackage{subcaption}
\usepackage{caption}
\usepackage[official]{eurosym}
\usepackage{multirow}
\usepackage{float}
\usepackage{verbatim}
\usepgfplotslibrary{units}
\usepackage{units}
\usepackage{mathtools}
\usetikzlibrary{arrows,shapes,positioning,matrix}
\usepackage{smartdiagram}
\usepackage{booktabs}
\usepackage[b]{esvect}
\usepackage{algorithm}
\usepackage{algorithmic}
\usepackage{authblk}
\usepackage{xr-hyper}
\usepackage[colorlinks=false, pdfborder={0 0 0}]{hyperref}
\usepackage{tikz-network}
\usetikzlibrary{positioning}
\usepackage[flushleft]{threeparttable}
\usepackage{units}

\usepackage[babel,german=quotes]{csquotes}
\usepackage[style=authoryear, date=year, doi=false,isbn=false, dashed=false, firstinits=true,maxbibnames=99,uniquelist=false,maxcitenames=3,backend=bibtex]{biblatex}
\bibliography{Literature}
\renewbibmacro{in:}{}
\renewbibmacro*{volume+number+eid}{%
  \printfield{volume}
  \addnbthinspace
  \printfield{number}%
  \printfield{eid}}
\DeclareFieldFormat[article]{number}{\mkbibparens{#1}}

\setkomafont{sectioning}{\normalfont\normalcolor\bfseries}

\newtheoremstyle{1}
	{\topsep}
	{\topsep}
	{\itshape}
	{}
	{\bfseries}
	{.}
	{ }
	{\thmname{#1}\thmnumber{ #2}\thmnote{ (#3)}}

\newtheoremstyle{2}
	{\topsep}
	{\topsep}
	{\normalfont}
	{}
	{\bfseries}
	{.}
	{ }
	{\thmname{#1}\thmnumber{ #2}\thmnote{ (#3)}}

\theoremstyle{1}
\newtheorem{theorem}{Theorem}[section]

\newtheorem{remark}[theorem]{Remark}

 \numberwithin{equation}{section}
 
\makeatletter
\newcases{mycases}{~}{%
  \hfil$\m@th\displaystyle{##}$}{$\m@th\displaystyle{##}$\hfil}{\lbrace}{.}
\makeatother

\makeatletter
\AtBeginDocument{%
  \expandafter\renewcommand\expandafter\subsection\expandafter{%
    \expandafter\@fb@secFB\subsection
  }%
}
\makeatother

\allowdisplaybreaks

\newcommand{\bbr}{\mathbb{R}}
\newcommand{\bbn}{\mathbb{N}}
\newcommand{\ncal}{\mathcal{N}}
\newcommand{\lcal}{\mathcal{L}}

\definecolor{red}{rgb}{1,0.1,0.2}

\title{Microscopic Traffic Models, Accidents, and Insurance Losses}
\author{Sojung Kim, Marcel Kleiber \& Stefan Weber}
\affil{Leibniz Universität Hannover}
\date{\today}

\begin{document}

\maketitle

\begin{abstract}
The paper develops a methodology to enable microscopic models of transportation systems to be accessible for a statistical study of traffic accidents. Our approach is intended to permit an understanding not only of historical losses, but also of incidents that may occur in altered, potential future systems. Through such a counterfactual analysis, it is possible, from an insurance, but also from an engineering perspective, to assess the impact of changes in the design of vehicles and transport systems in terms of their impact on road safety and functionality. 

Structurally, we characterize the total loss distribution approximatively as a mean-variance mixture. This also yields valuation procedures that can be used instead of Monte Carlo simulation. Specifically, we construct an implementation based on the open-source traffic simulator SUMO and illustrate the potential of the approach in counterfactual case studies.\\

\end{abstract}

\section{Introduction}

Every year, traffic accidents cause substantial damage, both property damage and injuries and deaths. For example, nearly 43,000 people died in road traffic accidents in the USA in 2021 (\textcite{NHTSA2022}). The frequency and severity of these accidents depends on the driving behavior of vehicles, on the one hand, and on the characteristics of traffic systems themselves, on the other. Improvements in road safety are achieved, for example, by reducing serious injuries in accidents through the design of vehicles, by car body design, airbags, seatbelts, etc. However, the frequency and type of accidents can also be influenced by modifying the transportation system itself and by changes in driving behavior. From a higher-level perspective, at least two dimensions are central, and we will examine them in this paper:
\begin{enumerate}
\item \emph{Engineering.} From an engineering perspective, the focus is on the good design of vehicles and traffic systems, combining functionality and safety. Instruments in this respect include traffic rules and their implementation, the layout of streets, and innovation in vehicle technology such as advanced driver-assistance systems or autonomous driving software. Improvements of this type may reduce the number of accidents and their severity, but cannot completely prevent accidents. 
\item \emph{Insurance.} Residual risks remain, and accidents cannot be completely prevented. However, at least in financial terms, the associated losses can be covered by insurance contracts. The role of actuaries is to develop adequate contract structures, calculate correct premiums, and implement quantitative risk management in insurance firms. These tasks require the modeling and analysis of probability distributions of accident frequencies, corresponding damages, and insurance losses.
\end{enumerate}
The objective of this paper is to develop a methodology to enable microscopic models of transportation systems to be accessible for a statistical study of traffic accidents. Our approach is intended to permit an understanding not only of historical losses, but also of incidents that may occur in altered, potential future systems. Through this, it is possible, from both an engineering and insurance perspective, to assess changes in the design of vehicles (e.g., the driving behavior of autonomous vehicles) and transport systems in terms of their impact on functionality and road safety. This is in stark contrast to simply considering aggregate data, as is typical in auto insurance. Instead our model can simulate traffic in counterfactual situations that mimic modified or future scenarios, and losses can be generated conditionally on local traffic conditions.

The conventional approach is as follows: To understand current traffic events or structural relationships in the past, historical data are used. Historical data can also be applied to test whether a model framework is appropriate in principle to describe traffic systems realistically. These data also constitute the essential basis for the specific pricing of insurance contracts in practice.  

But how can we examine risks associated with new technologies and with novel future strategies for traffic systems? Consider autonomous vehicles, for example: due to their altered driving behavior, these will reshape existing traffic patterns, and in turn, accident occurrences and associated losses. Insurance companies presumably will have to adapt their business models as well; in the future, premiums for auto insurance may depend upon the driving configuration of the vehicle rather than the risk profile of the driver. Real-time insurance rates for individual trips could also become important, with prices depending on the current traffic situation.

In order to investigate future developments, we are suggesting to devise simulation tools in analogy to digital twins of real transport systems, which allow counterfactual case studies of possible future transport systems.
The digital twin paradigm refers to the triad of a ``physical entity, a virtual counterpart, and the data connections in between'' (\textcite{Jones2020}). In our application, the physical entity is the (future) real-world transportation system for which data on losses are not yet available. Its virtual counterpart is the model we are building. Counterfactual case studies can be used to generate data, evaluate future driving technologies and their impact on accident losses. Based on the results, newly developed concepts (e.g., modified traffic rules, novel insurance coverage and their insurance premiums, etc.) can be adapted in the real world. The concept of the digital twin makes it possible to experiment with technologies and policies, and their effects on accident damage without having to implement risky tests in reality.

Methodologically, this paper combines existing microscopic traffic models with probabilistic tools from actuarial science and quantitative risk management to study accident damage and insurance losses in the context of simulations. In contrast to standard insurance practice, we model losses conditionally on specific traffic situations. These traffic situations are generated within a flexible micro-simulation model. As a specific example, we use the well-established traffic simulator SUMO (\textcite{Lopez2018}) to illustrate how traffic systems could be realistically modeled.  

We extend microscopic traffic models to include random accidents and corresponding losses. The losses are modeled as random variables whose distributions depend on microscopic data. Since insurance contracts typically cover annual periods, we set up a model for aggregate losses over a one-year time horizon. We also show that aggregate losses can be approximated by a mean-variance mixture of Gaussian distributions. This provides an alternative perspective on the distribution of the aggregate loss and a second method of evaluation besides crude Monte Carlo sampling. For certain insurance contracts, we improve the accuracy of the approximation-based valuation by using a correction term. This was originally developed by \textcite{kj2009} for the efficient pricing of complex financial instruments, there in the context of a classical Gaussian approximation.

Our digital twin approach enables a comprehensive analysis of risk in transportation systems: We study the impact of fleet sizes and their driving configurations on system efficiency and insurance prices. System efficiency is measured using traditional traffic statistics based on local traffic counts such as traffic flow, average speed, and density. Insurance claims are examined in terms of their probability distributions and selected statistical functionals. The innovation of our approach lies in the fact that counterfactual traffic scenarios can be considered which allow to test insurance solutions beyond historical data. This could also include online insurance products that depend on specific traffic situations.

The main contributions of this paper are:
\begin{enumerate}
\item We develop a powerful methodological framework to generate accident data based on microscopic traffic models in analogy to the concept of digital twins.
\item Specifically, we construct an implementation based on the state-of-the-art open-source traffic simulator SUMO and illustrate the potential of the approach in comprehensive case studies.
\item Structurally, we characterize the total loss distribution approximatively as a mean-variance mixture. This also yields alternative valuation procedures. These results hold for general microscopic traffic models; SUMO is used in case studies to illustrate the findings.
\item Based on Stein's method, we obtain a correction term in the valuation, derived from the results of \textcite{kj2009}, which enables surprisingly accurate pricing of insurance contracts.
\end{enumerate}

\subsection{Outline} 

The paper is organized as follows. Section~\ref{sec:lit} discusses related contributions in the literature. Section \ref{sec:model} presents the microscopic traffic model that captures also accidents. Section \ref{eval} discusses the evaluation of the losses. Case studies are presented in Section \ref{sec:applic}. Section \ref{sec:conclude} concludes and discusses further research challenges. The supplementary material contains in Appendix \ref{sec:sampling} details on the implemented sampling procedure; further simulation results, not presented in Section \ref{sec:applic}, are documented in Appendix \ref{sec:tables}. 

\subsection{Literature}\label{sec:lit}

Our paper combines microscopic traffic models with probabilistic tools from actuarial science and quantitative risk management to study risks in traffic systems. The literature can be classified along two dimensions, the engineering perspective and the actuarial perspective.  

\textbf{The Engineering Perspective.} An important field of operations research is the analysis and optimization of road traffic systems (see, e.g., \textcite{Gazis2002}) with respect to their efficiency. Traffic models are indispensable tools for this purpose: Macroscopic models are based on the functional relationships between macroscopic features such as traffic flow, traffic density, and average speed. These models allow the study of issues such as the efficient routing of vehicles under different constraints (see, e.g., \textcite{Acemoglu2018}, \textcite{ColiniBaldeschi2020}). Stochastics can be used to extend such risk considerations in terms of uncertain travel times (e.g., \textcite{Nikolova2014}). In the context of various applications of transportation systems, tailored stochastic models provide suitable analytical tools; the extensive literature includes, for example, the efficient routing of ambulances (\textcite{Maxwell2010}) or the allocation of capacity in bike-sharing systems (\textcite{Freund2022}).

To model transportation systems at a level of higher granularity, microscopic traffic models are used (see, e.g., \textcite{Helbing2001}). Their simulation, i.e., the computation of trajectories from accelerations, is computationally more demanding. There are established software solutions that facilitate the application of microscopic models. In this work, we use in the context of illustrative case studies the simulation engine SUMO (see the Section~\ref{sec:SUMO} for an overview). Examples of competing microscopic traffic simulators include VISSIM (\textcite{Fellendorf2010}) and Aimsun (\textcite{Casas2010}). While SUMO is open source software, these competitors are commercial. Software packages such as SUMO are also important in the context of testing and validating new technological developments such as autonomous vehicles. This is, for example, discussed in \textcite{Schwarz2022}, \textcite{Szalai2020}, and \textcite{Kusari2022}.
To deploy autonomous vehicles in the real world, lengthy and expensive testing phases are required. Acceleration strategies are being developed to shorten these times (e.g., \textcite{Zhao2018}). These approaches rely on importance sampling techniques to overcome the rare-event nature of safety-critical situations. \textcite{Arief2018} develop simulation-based testing methodologies in order to analyze autonomous vehicles in relevant scenarios that are constructed using collected data.   \textcite{Norden2019} create a framework for the black-box assessment of the safety of autonomous vehicles. They apply their framework on a commercial autonomous vehicle system. Our work focuses on aggregate losses over relatively long time horizons. By considering one-year losses via a conditional loss modeling approach, we bypass the problem of simulating rare events.
Further literature on microscopic traffic models, their calibration, and applications for traffic safety is reviewed in the online appendix in Appendix \ref{sec:appendixReview}.

\textbf{The Actuarial Perspective.} The ambitious goal of achieving maximum efficiency and complete safety through engineering design cannot be realized in reality; accidents can never be completely excluded, even if residual risks can be kept very small. Insurance is an instrument to deal with the residual risk of infrequent losses, cf.  \textcite{McNeil2015} and \textcite{Wuthrich2013}.

The premiums of motor insurance contracts are traditionally based on historical claims data collected by insurance companies. Insurance premiums are calculated based on individual characteristics of the driver (age, driving experience, etc.) and the vehicle (type, location, etc.). These tariffs are often complemented by bonus-malus schemes (see, e.g., \textcite{Denuit2007}, \textcite{Lemaire2015},  \textcite{Afonso2017}) to incentivize more careful driving and prevent insurance fraud.

Novel pricing approaches use telematics technology (see, e.g., \textcite{Husnjak2015} for an overview). This involves collecting GPS data from vehicles, which can be analyzed and classified. Machine learning techniques are suitable to process these large amounts of data. We refer to \textcite{Gao2022} for a methodological overview. \textcite{Verbelen_2018},  \textcite{Corradin2021}, \textcite{So2021}, \textcite{Henckaerts2022} discuss telematics pricing and usage-based auto insurance products.

Our approach can be understood as complementary to telematics pricing: Instead of analyzing driving data to determine the driving behavior of individuals, we model the behavior of vehicles as a \emph{driving configuration} and subsequently generate driving data and insurance claims. Our approach is in particular suitable, if novel technologies are studied in counterfactual situations, e.g., autonomous vehicles. To our knowledge, there is no other work that develops a microfounded model of traffic accidents that can be leveraged to study insurance pricing. 

\section{Model Components}\label{sec:model}

Our microfounded simulation model for investigating accident losses is based on two components:
\begin{enumerate}
\item At its core is a deterministic microscopic traffic model that realistically characterizes the motion of vehicles in a traffic system typically represented by a system of ordinary differential equations. The particular strength of the simulation-based approach is that also counterfactual situations can be modeled -- capturing, for example, the consequences of technological innovation.
\item This microscopic traffic model is extended to include the possibility of random accidents. At random accident times, local traffic data are observed which characterize the probability distribution of the occurring losses. In our specific implementation of this general conceptual approach, the SUMO microscopic traffic simulator is used, but our theoretical results hold for any microscopic traffic model that satisfies the properties described in the next section.
\end{enumerate}

\subsection{Microscopic Traffic Networks}

We consider a road network that is typically embedded into a two-dimensional area $A\subseteq\bbr^2$. The network may consist of roads, junctions, roundabouts, intersections, highways, etc. on which vehicles move. The collection of all vehicles in the network is denoted by $\mathcal{M}$. Each vehicle $i\in\mathcal{M}$ is assigned an origin-destination pair $(O^i,D^i)\in A$. 

We consider a fixed time horizon $T>0$. Vehicles move over time from their origin to their destination on a (potentially changing) path. We denote by $x^i(t)$ the position of vehicle $i$ at time $t\in[0,T]$, by  $v^i(t)=\frac{\mathrm{d}}{\mathrm{d}t}x^i(t)$ their velocity, and by $a^i(t)=\frac{\mathrm{d}}{\mathrm{d}t}v^i(t)$ their acceleration. We make the implicit assumption that vehicles are located in $O^i$ until some release time and remain in $D^i$ once reached. Thus, we let $\mathcal{M}(t)=\left\{i\in\mathcal{M}\colon x^i(t)\notin\{O^i,D^i\}\right\}$ be those vehicles which are currently \emph{inside} the network, i.e., they have left their origin but not reached their destination, yet. If we only consider vehicles which belong to a certain group of vehicles (also called a \emph{fleet}) $\Phi\subseteq\mathcal{M}$, we will write $\mathcal{M}^\Phi(t)$. This could, for example, be a fleet of vehicles with the same driving characteristics.

At the core of many microscopic traffic models are \emph{car-following models.} Prominent examples include the Intelligent Driver Model (\textcite{Treiber2000}), the Optimal Velocity Model (\textcite{Bando1994} and \textcite{Bando1995}), and the Krau{\ss} model (\textcite{Kraus1998}). Car-following models determine the acceleration behavior of an individual vehicle $i$ along its path on the basis of information on the positions and velocities of the vehicles, typically in a neighborhood of $i$, and the properties of the system. Often, only the preceding vehicle on the road is relevant, and the acceleration of $i$ is constructed such that vehicles move forward while maintaining a minimal distance. Through specific choices more complex traffic scenarios (e.g., intersections, overtaking) can still be represented in such a manner. Mathematically, car-following models correspond to systems of coupled ordinary differential equations.

\textbf{Traffic State.} We denote by $\gamma(t)=\left(x^i(t),v^i(t),a^i(t)\right)_{i\in\mathcal{M}}$ the state of the traffic system at time $t$. It records the position, velocity, and acceleration of any vehicle. The evolution of the traffic system over time is depicted by the (high-dimensional) trajectory $t\mapsto \gamma(t)$.

Macroscopic traffic statistics aggregate these microscopic data. Typical examples include traffic flow (number of vehicles that pass a certain point per time unit), traffic density (number of vehicles per length unit), and average speed. These measures quantify the performance of traffic systems.

\textbf{Local Traffic Conditions.} In order to model the occurrence of accidents depending on local traffic conditions, we partition $A$ in regions. More precisely, we partition $A$ into a finite number of disjoint sets $A_r\subseteq A$ such that $ A = \bigcup_{r=1}^R A_r$ and $R \in\mathbb{N}$. We call the elements $A_r$ of the partition a \emph{traffic module}. 

We let $\mathcal{M}_r(t)=\left\{i\in\mathcal{M}(t)\colon x^i(t)\in A_r\right\} \subseteq \mathcal{M}(t)$ denote those vehicles that are in $A_r$ at time $t$ (with $\mathcal{M}^\Phi_r(t)$ defined in analogy). The local traffic state of the module is $\gamma_r(t) = ( x^i(t), v^i(t), a^i(t))_{i \in \mathcal{M}_r(t)}$. Key local traffic characteristics (density, flow, speed, etc.) can then be expressed as functions of $\gamma_r(t)$ and its evolution over small time windows.

\subsection{Microscopic Traffic Model With Accident Losses}\label{mtmwal}

So far, the evolution of the traffic system is a deterministic function of time. The advantage of microscopic models is that they enable a detailed simulation of traffic systems. The driving behavior of the vehicles can be varied, likewise their number and paths, road conditions, etc.~in order to generate many different scenarios. Such models provide a detailed picture, similar to digital twins of reality, and can be used to analyze potential future traffic systems or to understand the impact of new technologies. 

We consider a finite collection of different traffic scenarios $\gamma^k := (\gamma^k(t))_{t\in [0,T]}$ with $k \in \{1,2, \dots, K  \} $ for a short time horizon $T>0$. The aim is to analyze characteristics of traffic over the long time horizon $N  T$, e.g., one year, for some large $N\in \bbn$; this is modeled by a finite sequence of traffic scenarios $(k_1, k_2, \dots, k_N) \in \{1,2, \dots, K  \}^N $. The $N$ subintervals of length $T$ are called time buckets. We will be interested in quantities aggregated or averaged over the whole time horizon $NT$. Examples include the average traffic flow, the total number of accidents, the aggregate losses due to accidents, etc. These quantities do not depend on the order of the traffic scenarios during this time period, but only on their number of occurrences. 

We denote by $\mu^k$ the number of occurrences of scenario $k$ divided by $N$, i.e., the relative frequency of this traffic scenario over the considered time horizon $N T$. The vector $\mu= (\mu^1, \mu^2, \dots, \mu^K)^\top$ lies in the simplex $\Delta^{K-1} = \{x \in \bbr^K_+: \sum_{k=1}^K x^k =1 \}$. We assume that $\mu$ is not deterministic, but a random variable. This is to account for the fact that the relative frequencies of traffic scenarios fluctuate over different years due to varying weather conditions, random changes in traffic demand, or other factors. From a mathematical point of view, this construction leads to a mixture model with exogenous factor $\mu$.

\textbf{Accident Occurences.} We now introduce our traffic accident model that will permit an analysis of aggregate losses and corresponding insurance contracts. The likelihood of the occurrence of an accident is modeled as a function of the traffic scenario. We consider two specifications, a Binomial and a Poisson model.
\begin{enumerate}
\item \emph{Binomial Model.} Accidents are rare events. For a given traffic scenario $k$, we assume that the probability  $p^k$ of an accident is close to zero. This probability may, of course, depend on the evolution of the traffic scenario, i.e., on the path $t\mapsto \gamma^k(t)$, and we will discuss concrete specifications later. Given a realization of $\mu$, accidents are assumed to be independent across time buckets. This implies that traffic scenario $k$ occurs for $N  \mu^k$ time buckets corresponding to a duration of $N T  \mu^k$, and the number of accidents $C^k$ during this period has a conditional Binomial distribution with parameters $p^k$ and $N  \mu^k$:
$$C^k \mid \mu  \sim {\rm Bin} (p^k, N  \mu^k ).$$
\item \emph{Poisson Model.} An alternative model assumes that accidents occur at random times with a distribution governed by an intensity $\lambda^k/T$ that depends on the traffic scenario $k$. More specifically, the number of accidents $C^k$ during the period governed by scenario $k$ of duration $N T  \mu^k$ is conditionally Poisson distributed with parameter $\lambda^k N \mu^k$:
$$C^k  \mid \mu \sim {\rm Poiss} (\lambda^k N \mu^k).$$
\end{enumerate}

\textbf{Accident Losses.} Loss sizes conditional on the occurrence of accidents are assumed to be independent across traffic scenarios and across time buckets. We assume that the conditional loss distribution with conditional distribution function $F^k$ depends only on the traffic scenario $k$. We will discuss examples below. Random total losses over the considered time horizon $NT$ are equal to 
$$L\; = \;   \sum_{k=1}^K \; \sum_{c=1}^{C^k}  \;  X^k_c   $$
where the random variables $X^k_c$, $k=1,2, \dots, K$, $c\in \bbn$, are independent and $X^k_c \sim F^k$, $c\in\bbn$, for any $k$. 

\textbf{Concrete Specifications.} Microscopic traffic models are experimental environments that allow to simulate the behavior of systems where no real data are yet available. Traffic planning can be supported by such models, and the impact of new technologies can be tested in a counterfactual analysis. Here, we specify the general principles how accident occurrences and losses can be based on microscopic traffic models. An implementation will in this paper be based on SUMO, see Section \ref{sec:SUMO}, but could also rely on any other suitable traffic model.

Initially, $K$ traffic scenarios need to be selected as a basis for the model. While running any deterministic traffic scenario $k$ over the  time window $[0,T]$, information can be extracted about the traffic states $\gamma_r^k$ in each module $r\in \{1,2, \dots, R\}$. In SUMO typically not complete data on the whole paths are extracted, but only selected information at loop detectors in the network that are part of the implementation.

In reality, the likelihood of accidents typically increases with higher traffic density and higher velocities, ceteris paribus. Also the distribution of losses is influenced by quantities of this type. Examples are described in Section~\ref{sec:applic}. This allows a computation of $p^k$ and $\lambda^k$ as a function of the data. Using the data associated with the modules, we may specify probabilities and intensities for the modules such that $p^k= \sum_{r=1}^R p^k_r$ and $\lambda^k = \sum_{r=1}^R \lambda^k_r$. In the Binomial model, $p^k_r/p^k$ is the conditional probability that the accident is in module $r$ given that an accident occurs. In the Poisson model, the intensities $\lambda^k_r$, $r=1,2, \dots, R$, determine the accident times for each module. The resulting sequence of random times in the whole traffic system possesses the intensity $\lambda^k$. Conversely, if we first simulate random times with intensity $\lambda^k$ and then randomly choose a corresponding module with probability $\lambda^k_r/\lambda^k$ in a second step, the random times associated with each module $r$ possess intensity $\lambda^k_r$. Both procedures produce a number of accidents $C^k$ that occur during the period governed by scenario $k$.

The distributions of losses given a single random event will be chosen as follows. For each traffic scenario, we consider a collection of distribution functions $(F^{k,\psi})_{\psi\in \Psi}$ where $\psi$ corresponds to data that may be extracted from the traffic simulation. In order to do so, we uniformly simulate a random time in $[0,T]$ and extract at this time the data from scenario $k$ that determine $\psi$. The resulting distribution $F^k$ is a mixture of the distributions $(F^{k,\psi})_{\psi\in \Psi}$. The mixing distribution is derived on the basis of the traffic data of scenario $k$ that are generated from our microscopic traffic  model.

\textbf{Insurance Contracts \& Statistical Functionals.} The microscopic traffic model with accidents will be the basis for the generation of aggregate losses. We study statistical functionals and insurance contracts. The analysis will be based on Monte Carlo simulations, but we also compare approximation techniques that we describe in Section~\ref{eval}. 

The focus is on functions of aggregate losses $L$. Letting $h\colon \bbr \to \bbr $ be an increasing function, we analyze $h(L)$. In particular, we investigate the following functions corresponding to three types of insurance coverage: $h(x)=x$ (full coverage), $h(x)=\max(x-\theta,0)$, $\theta\geq 0$ (constant deductible), $h (x) =\min(x,\theta)$, $\theta\geq 0$ (stop loss). In each case, we evaluate various statistical functionals:
\begin{enumerate}
\item \emph{Expectation.} $\mathbb{E}(h(L))$, 
\item \emph{Variance.} $\mathrm{Var}(h(L))=\mathbb{E}((h(L))^2))-\mathbb{E}(h(L))^2$,
\item \emph{Skewness.} $\varsigma_{h(L)}=\frac{\mathbb{E}\left[(h(L)-\mathbb{E}(h(L)))^3\right]}{(\mathrm{Var}(h(L)))^{3/2}}$,
\item \emph{Value-at-Risk.} $\mathrm{VaR}_{p}(h(L))=\inf \{x\in \mathbb{R}\colon P(h(L)\leq x)\geq p\}$,
\item \emph{Expected Shortfall.} $\mathrm{ES}_{p}(h(L))=\frac{1}{1-p}\int_p^1 \mathrm{VaR}_q(h(L))\,\mathrm{d}q$.
\end{enumerate}
These functionals allow also the computation of insurance premiums on the basis of premium principles such as the expectation principle, the variance principle, or the standard deviation principle.

\subsection{Traffic Scenarios in SUMO}\label{sec:SUMO}

\subsubsection{A Brief Overview}

A state-of-the-art open source software that allows us to generate traffic scenarios is SUMO, ``\textbf{S}imulation of \textbf{U}rban \textbf{MO}bility''. A reference publication on SUMO is \textcite{Lopez2018}; in addition a detailed user documentation can be found online (see \url{sumo.dlr.de/docs/index.html}). Freely available since 2001, SUMO was originally developed by the German Aerospace Center and extended by an active research community. It allows for a plethora of modeling choices at different levels and has been successfully applied to tackle many important research questions addressing (see \url{eclipse.org/sumo/about/}), e.g., traffic light optimization, routing, traffic forecasting, and autonomous driving.

\begin{figure}
\centering
\includegraphics[scale=0.5]{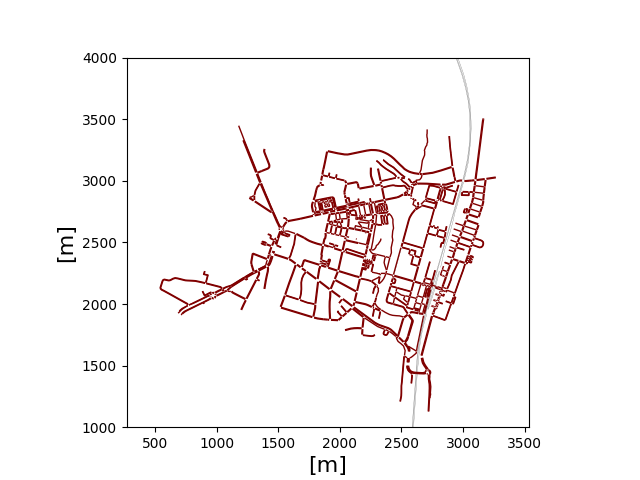}

\caption{SUMO network of Wildau.}
\label{fig:SUMOWildau}
\end{figure}

In the following, we give a short overview. At its core, SUMO is a software which generates a traffic scenario $\gamma=(\gamma(t))_{t\in[0,T]}$ from a given set of input files: 
\begin{enumerate}
\item \emph{Network File.} In SUMO, a traffic network is described by a directed graph whose nodes represent intersections and edges roads. All nodes and edges have attributes including positions, shapes, speed limits, traffic regulation, etc. As an example, the city of Wildau is represented as a SUMO network in Figure \ref{fig:SUMOWildau}. 
\item \emph{Route File.} Vehicles are generated on the basis of \emph{traffic demands} between origins and destinations. Routes can either be defined for each vehicle as a \emph{trip} or as recurring \emph{flows} along a specific path. If only origin and destination are provided, the corresponding route is computed when the vehicle enters the system.  The problem of allocating traffic demand to routes in a network is referred to as the traffic assignment problem. A standard reference is \textcite{Patriksson2015}.

The \emph{vehicle type} determines the microscopic characteristics of the vehicle such as the governing car-following model, driving parameters (e.g., maximal speed, maximal acceleration, time headway), size, color, etc. By default, vehicles are passenger cars. Other modes, such as pedestrian, bicycle, or truck can also be selected.
\item \emph{Additional Files.} Further components are specified in additional files. An important example are induction loop detectors. These  collect time series data on aggregate traffic statistics by counting the vehicles which pass a certain position during a short time interval.
\end{enumerate}

The collection of input files determines the traffic evolution, also called the \emph{SUMO scenario}. The computation can be executed either as a command line application or with a GUI that visualizes the movement of the vehicles through the network over time.

\textbf{Data Extraction.} One particularly appealing extension of SUMO is the ``\textbf{Tra}ffic \textbf{C}ontrol \textbf{I}nterface'', TraCI (see \textcite{Wegener2008}). TraCI provides online access to the microscopic traffic simulation and permits, at each time step, through a comprehensive list of commands (a detailed description can be found at \url{sumo.dlr.de/docs/TraCI.html}) to retrieve data and to change the states of objects such as vehicles, roads, traffic lights, etc. Available in standard programming languages (our case studies are based on the Python implementation), TraCI yields easy access to SUMO without the need to modify the underlying code. We use TraCI to extract microscopic data on positions, velocities, and accelerations of randomly selected vehicles.

\subsubsection{Generation of Traffic Scenarios}\label{sec:SumoScenario}

To represent traffic in a given area over a longer time horizon (e.g., $NT=\unit[1]{year}$), we generate a diverse collection of traffic scenarios $\gamma^1,\dots,\gamma^K$ of duration $T$ in SUMO by varying the input files. Traffic over a longer time horizon is represented by a random composition of these traffic scenarios. Our general construction has already been discussed in Section~\ref{mtmwal}.

\textbf{SUMO Scenario.} We describe the key steps to set up a SUMO scenario and point out specific references to the SUMO documentation in the online appendix in Section \ref{sec:sumoReferences}. SUMO provides tools that facilitate the creation of input files, e.g., the graphical network editor \emph{netedit} that visualizes a SUMO scenario and allows to modify its properties. In practice, network files are typically imported from other data sources. For example, one can build a real-world traffic network in SUMO from OpenStreetMap data by selecting an area from a map.
The route file specifies the trips of the vehicles and the definition of the general vehicle types with their microscopic characteristics. First, there are several options to generate trips in SUMO. These can be obtained from empirical data in the form of traffic counts, imported to SUMO as origin-destination matrices, or modeled via ad-hoc choices, e.g., using \emph{netedit}. Second, each vehicle is associated to a vehicle type specified on the basis of a comprehensive list of attributes. The corresponding values can be set manually in the route file or accessed and modified via \emph{netedit}.
In the absence of detailed traffic data for calibration, SUMO offers an \emph{activity-based demand generation} which deduces traffic demand from general assumptions on the structure of the population (inhabitants, households, etc.) in the considered area. The tool \emph{activitygen} automates the process and produces an artificial route file.

SUMO admits a large variety of modeling choices. Tailored to the needs of the modeler, a highly detailed SUMO scenario can be constructed. Our case studies will be based on publicly available SUMO scenarios; these consist of network and route files which are calibrated to real-world cities.

\textbf{Varying Traffic Conditions.} The input files need to be constructed in such a way that they reflect varying traffic conditions over longer time periods. This includes weather conditions, variation of traffic demand, and other factors. 

\textcite{Maze2006} review empirical studies on the impact of adverse \textit{weather conditions} on traffic. These may induce i) lower traffic demand, ii) higher risk of accidents, and iii) modified driving behavior. Based on empirical findings, \textcite{Phanse2022} implement reduced velocities due to rainfall. In \textcite{Weber2019}, the idea of introducing into SUMO a friction parameter per road is discussed. Traffic scenarios under adverse weather conditions can be captured by suitable driving parameters in the route file (e.g., by variation of maximal speed, maximal acceleration, etc.), and this may be combined with a weather-dependent model of the occurrence and the severity of accidents.

\textit{Traffic demand} is traditionally estimated from traffic counts. We refer to  \textcite{Bera2011} for an overview. With increasing data availability, the estimation can be enhanced by floating car data (cf., e.g., \textcite{Nigro2018}), i.e., data generated from vehicles over time as they are driving. Traffic demand varies over time, but patterns reoccur over longer time horizons (see, e.g., \textcite{Soriguera2012}). Demand depends on the considered traffic network. Weekdays differ from days on the weekend; peaks in demand occur at common commute times. Rush hours are spatio-temporal phenomena that can be analyzed in detail (see, e.g., \textcite{Xia2018}). 

To reflect the heterogeneity of traffic scenarios, two options are available in SUMO: i) A variety of route files is generated that is consistent with the desired modeling granularity. This process can  be automatized via an additional program, a \emph{route file generator}, that produces route files with the desired characteristics. ii) Another option is to select a medium time horizon (e.g., $T_\mathrm{SUMO}=\unit[24]{h}$) with a corresponding route file that depicts varying traffic demand over time. From the generated SUMO scenario, a selection of small time horizon scenarios (e.g., $T=\unit[1]{min}$) can be efficiently generated by utilizing SUMO's option to save the state of the running simulation at a priori specified times and load these later.

Besides weather and traffic demand, many other factors influence the traffic dynamics. \textcite{Wagner2016} discusses the representation of autonomous vehicles in SUMO. \textcite{Luecken2019} utilize SUMO to study control transition, i.e., selected safety critical situations where the human driver needs to take over control from an autonomously driving vehicle. \textcite{Pagany2020} study the impact of wildlife on traffic, an issue that is also relevant in the context of traffic accidents.

\section{Evaluation Methods}\label{eval}

The accident losses $L$ can be simulated using Monte Carlo methods. The simulations may be used to estimate the value of statistical functionals and to price insurance products. We will briefly describe the Monte Carlo methods. In addition, on the basis of the Binomial model, we construct a Gaussian approximation to $\mathbb{E}(h(L))$ where the function $h$ corresponds to the three types of insurance coverage that we consider: full coverage, constant deductible, and stop loss. This allows a numerical evaluation similar to \textcite{fpw2008} and \textcite{kj2009}. The latter paper provides a correction term derived by Stein's method that we will exploit in our application.

\subsection{Monte Carlo Methods}

The Monte Carlo simulation of $L$ requires sampling the number of accidents $C^k$ in either the Binomial or Poisson model and sampling the independent conditional losses $X^k_c$, $c\in \bbn$, for each traffic scenario $k=1,2, \dots, K$ from the corresponding distribution $F^k$.  These tasks can be performed separately, but require both a prior evaluation of the microscopic traffic model. 

\begin{enumerate} 
\item \emph{Prior Evaluation of Traffic Model.} For each traffic scenario $k$ a single run delivers data that are the basis for a computation of, respectively, the accidents probabilities $p^k$ and intensities $\lambda^k$ as well as the corresponding values $p^k_r$ and $\lambda^k_r$ on the level of the modules $r=1,2, \dots, R$.
\item \emph{Number of Accidents.} Sampling from $\mu$ and using the results of the prior evaluation allows to sample the number of accidents $C^k$ for each traffic scenario $k$ in both the Binomial and Poisson model.
\item \emph{Conditional Accident Losses.} Based on the precomputed values of the accident probabilities and the accident intensities, respectively, we simulate for each traffic scenario $k$ the random locations and times of accidents. These  data can be stored. For these locations and times, traffic data $\psi$ are extracted from an additional single run of traffic scenario $k$. Given $\psi$, the losses are generated according to conditional loss distributions $F^{k, \psi}$ as described in Section~\ref{mtmwal}. Details of the implementation are explained in Section~\ref{sec:applic}.
\end{enumerate}

These Monte Carlo methods can be flexibly applied to all considered functionals. In the special cases of expectation and variance, Wald's equation can simplify the computation, since the losses $L$ are given in the form of a collective model. 

\subsection{Gaussian Approximation}\label{sec:approximation}

Another way to compute $\mathbb{E}(h(L))$ for the considered types of insurance coverage is a Gaussian approximation, possibly improved by a correction term. A Gaussian approximation can easily be motivated within the Binomial model. For $N$ sufficiently large, the distribution of $L$ given $\mu$ is approximately normal, implying that $L$ is a mean-variance-mixture of Gaussian distributions. This is an important structural insight from this approximation.

In this section, we condition on $\mu$, i.e., suppose that $\mu$ is fixed and given. The general results for $\mu$ random are then a corollary by considering suitable mixtures according to the distribution of $\mu$.  The total random losses in the Binomial model can be rewritten as
$$L\; = \;   \sum_{k=1}^K \; \sum_{c=1}^{N\mu^k}  \;  \mathbf{1}^k_c \cdot X^k_c   $$
where the random variables $\mathbf{1}^k_c$, $X^k_c$, $k=1,2, \dots, K$, $c\in \bbn$, are independent, $X^k_c \sim F^k$, and $\mathbf{1}^k_c$ are Bernoulli random variables taking the value $1$ with probability $p^k$ and the value $0$ otherwise, $c\in\bbn$, for any $k$. Setting 
$$Y^k_c \; := \; \mathbf{1}^k_c \cdot X^k_c, \quad m^k \; := \; \mathbb{E}(Y^k_c), \quad {\left(\sigma^k\right)}^2 \; := \;  \mathbb{E}\left([Y^k_c - m^k ]^2\right), \quad {\left(\zeta^k\right)}^3 := \; \;  \mathbb{E}\left([Y^k_c - m^k ]^3\right),$$
$k=1,2, \dots, K$, $c\in \bbn$, a classical normal approximation of $L$ is $\sum_{k=1}^K N\mu^k m^k + Z $ with 
$$Z \quad \sim \quad \ncal \left(  0 , \; \sum_{k=1}^K  N\mu^k {\left(\sigma^k\right)}^2   \right).$$

\begin{remark}
The estimation of $m^k$, ${\left(\sigma^k\right)}^2$, and ${\left(\zeta^k\right)}^3$ requires the simulation of the random variables $X^k_c$, $c\in \bbn$. The independent terms  $\mathbf{1}^k_c$, $c\in\bbn$, factor out, are idempotent and have known expectation $p^k$.
\end{remark}

We focus on three types of insurance coverage, $h(x)=x$ (full coverage), $h(x)=\max(x-\theta,0)$, $\theta\geq 0$ (constant deductible), $h=\min(x,\theta)$, $\theta\geq 0$ (stop loss), and obtain an approximation $\mathbb{E}(h(Z))$ of $\mathbb{E}(h(L))$ in each of these cases.

On the basis of Stein's method (see \textcite{Chen2011} and \textcite{Ross2011} for an overview), \textcite{kj2009} suggest correction terms in order to improve the approximation, i.e., the approximation $\mathbb{E}\left(h\left(\sum_{k=1}^K N\mu^k m^k + Z\right)\right)$ is replaced by the corrected approximation  $$\mathbb{E}\left(h\left(\sum_{k=1}^K N\mu^k m^k + Z\right)\right)\; + \; C_h . $$ The correction $C_h$ depends on the degree of smoothness of the derivatives of the function $h$ and thus differs (see Theorem 3.1 and Proposition 3.6 in \textcite{kj2009}) for the three types of coverage. We define
$$d_1 \; = \;  \sum_{k=1}^K N\mu^k m^k,     \quad d_2 \;= \; \sum_{k=1}^K  N\mu^k {\left(\sigma^k\right)}^2 , \quad d_3 \; = \; \sum_{k=1}^K N\mu^k \left(\zeta^k\right)^3, \quad \tilde h (x) \; = \; h\left(   x +  d_1 \right) .$$ We obtain the following correction terms:
\begin{enumerate}
\item \emph{Full Coverage.}  In the case of full coverage, the correction term of \textcite{kj2009} disappears. In general, if $h$ is some Lipschitz function with bounded third derivative, the correction term equals $C_h \; = \; \frac{d_3}{2 d_2^2} \cdot \mathbb{E}\left( \left\{  \frac{Z^2}{3 d_2} - 1  \right\} Z \tilde h(Z)    \right).$
\item \emph{Constant Deductible.}  \tabto{5cm} $C_h \; = \; \frac{ (\theta - d_1)d_3 }{6 d_2} \cdot \frac{1}{\sqrt{2 \cdot \pi \cdot  d_2}} \cdot \exp \left\{ - \frac{(\theta - d_1)^2}{2 \cdot d_2}\right\}$
\item \emph{Stop Loss.} A stop loss $x\mapsto\min(x,\theta)$ can be written as the difference between full coverage $x\mapsto x$ and a constant deductible $x\mapsto \max(x-\theta,0)$. This implies that the correction term for a constant deductible appears with a negative sign in this case. 
\end{enumerate}

The advantage of the (corrected) Gaussian approximation in comparison to pure Monte Carlo is that, once the numbers $m^k$,  ${\left(\sigma^k\right)}^2$, and $\left(\zeta^k\right)^3$ have been computed for each traffic scenario $k=1,2, \dots, K$, no further data need to be stored or sampled in order to compute $\mathbb{E}(h(L))$.  The approximate representation of the distribution of $L$ as a mean-variance-mixture is a considerable simplification.

\section{Application}\label{sec:applic}

We illustrate the application of our microscopic traffic model with accidents on the basis of a publicly available SUMO scenario of a real city.

\subsection{SUMO Scenario \& Accident Data}\label{sec:sce}

Wildau is a small German city of approximately 10,000~inhabitants, located around 30~km south-east of the capital Berlin. A SUMO model of the city was developed within a study project by the Technical University of Applied Sciences Wildau and is publicly available (see \url{github.com/DLR-TS/sumo-scenarios/tree/main/Wildau}).

\textbf{SUMO Scenario.} The implemented road network is visualized in Figure \ref{fig:SUMOWildau}. It is specified using 646 nodes connected by 1,426 edges. The city itself is crossed by the railway; the tracks are represented by the gray line. Vehicles in the present scenario are calibrated from real traffic counts provided by local authorities, see also \textcite{Behrisch2022}. Such numbers of vehicles per time unit for certain positions are then assigned to different routes through the city. The original scenario has a duration of $\unit[7,010]{s}$. Empty in the beginning, vehicles enter the system with a peak of approximately $240$ vehicles that drive simultaneously. In total, $2502$ vehicles are generated.

In the following section, we describe in detail how we adjust this SUMO scenario to obtain suitable ingredients for our case studies. This yields a collection of traffic scenarios that allow us to compare the effects of different driving characteristics and fleet sizes on the total loss and related insurance premiums.

\textbf{Varying Traffic Conditions.} For different collections of model parameters representing different traffic systems we generate adjusted SUMO scenarios. For each choice of parameters we proceed as follows to produce $K=100$ traffic scenarios of length $T=\unit[60]{s}$. Traffic scenarios $k=1,2, \dots, 50$ correspond to selected time intervals from the SUMO scenario. Traffic scenarios $k= 51, 52, \dots, 100$  represent higher traffic volumes. They are generated by replacing the original route file by a route file that consists of two copies of the original route file. This simple procedure generates a larger amount of vehicles along the original paths. The traffic scenarios are again selected time intervals from the corresponding SUMO scenario.

To represent the full year, we set $N=365\cdot 24\cdot 60=525,600$. We need to specify the random vector $\mu=(\mu^1,\dots,\mu^K)^\top$ describing the number of occurrences of the individual traffic scenarios divided by $N$. For the purpose of illustration, we specifically assume that two probability measures $\nu_g$, $\nu_b$  are given on $\{1,2, \dots, K \}$ which approximately correspond to the relative frequencies of traffic scenarios in two prototypical years $y=g,b$. In addition, we suppose that the type $y$ of the current year is random where both values $g$ and $b$ have probability $1/2$. Given $y$, we generate $\mu=(\mu^1,\dots,\mu^K)^\top$ from a  multinomial distribution corresponding to $\nu_y$. That is, for all time buckets $n=1,2, \dots, N$, a traffic scenario $k$ is chosen  independently from the distribution $\nu_y$ on $\{1,2, \dots, K\}$. Dividing the number of occurrences of a scenario $k$ by $N$, one obtains its random relative frequency $\mu^k$ for any $k=1,2, \dots, K$. In our case study, the distribution  $\nu_g$ corresponds to lower traffic densities on average, while $\nu_b$ is associated with higher traffic densities, i.e., we set $$
\nu_g \; := \; \left\{
\begin{array}{ll}
1/75, &  k=1,2, \dots, 50,\\
1/150, & k =51,52, \dots, 100
\end{array}
\right., \quad \nu_b \; := \; \left\{
\begin{array}{ll}
1/150, &  k=1,2, \dots, 50,\\
1/75, & k =51,52, \dots, 100.
\end{array}
\right. 
$$

\textbf{Accident Data.} The German Accident Atlasprovided by \textcite{Unfallatlas2022} depicts the locations of all police-reported accidents \emph{involving personal damage} that occurred within one year. In 2020, within the modeled area of Wildau (approximately) 48 accidents were registered. There are also aggregate statistics for Germany for all police-reported accidents. In 2020, approximately $\unit[11.8]{\%}$ of all road accidents involved personal damage (see \textcite{Bundesamt2023}). We use an estimate of $\bar{c}_{\mathrm{year}}=48/\unit[11.8]{\%}\approx 407$ accidents for calibration purposes.

\subsection{Model Specification}\label{sec:speci}

Our goal is to analyze accident losses for a fleet $\Phi$ over the time horizon of one year.

\textbf{Fleet Definition.} In the Wildau scenario, vehicles are defined using repeated flows from origins to destinations. Passenger cars (next to trucks and the train) are defined via 90 different flows. A flow generates vehicles of a given type at a given position. From this position, they navigate to a specified destination.

In SUMO, a passenger car represents a certain vehicle type. Initially, all passenger cars belong to the same vehicle type and, consequently, have the same driving characteristics. To introduce a fleet $\Phi$ of vehicles whose driving characteristics we can vary, we define a new vehicle type $\Phi$ and construct corresponding new SUMO scenarios. Fixing a fraction $\rho^\Phi\in[0,1]$ of vehicles belonging to $\Phi$, we retain approximately $1-\rho^\Phi$ of the existing flow definitions and modify $\rho^\Phi$ of the flow definitions suitably in order to model the fleet. In our case studies, we consider $\rho^\Phi = \unit[10]{\%},~\unit[50]{\%},~\unit[90]{\%}$.

\textbf{Driving Configuration.} Vehicles in a fleet $\Phi$ are of the same type. Various characteristics can be varied in SUMO; we focus on \emph{maximal speed} $v_\mathrm{max}$, \emph{maximal acceleration} $a_\mathrm{max}>0$, and \emph{time headway} $\zeta>0$. The time headway is the distance which is kept to the preceding vehicle measured in time, i.e., a velocity weighted safety distance. We refer to a fixed selection of driving characteristics as a \emph{driving configuration}. 

In our case studies, we will vary the driving configuration for all vehicles in the fleet $\Phi$ and keep all other vehicles as originally introduced; we use the implementation of an Intelligent Driver Model without any speed deviation that does not include further random effects. A driving configuration of vehicles in fleet $\Phi$ is denoted by $\xi=(v_\mathrm{max},a_\mathrm{max},\zeta)$. Specifically, we consider :
\begin{align*}
\xi^{1a}&=(\unit[5]{m/s},~\unit[0.8]{m/s^2},~\unit[3.0]{s}),~\xi^{2a}=(\unit[10]{m/s},~\unit[0.8]{m/s^2},~\unit[2.0]{s}),~ \xi^{3a}=(\unit[15]{m/s},~\unit[0.8]{m/s^2},~\unit[1.0]{s}), \\
\xi^{1b}&=(\unit[5]{m/s},~\unit[2.6]{m/s^2},~\unit[3.0]{s}),~\xi^{2b}=(\unit[10]{m/s},~\unit[2.6]{m/s^2},~\unit[2.0]{s}),~ \xi^{3b}=(\unit[15]{m/s},~\unit[2.6]{m/s^2},~\unit[1.0]{s}).
\end{align*}
The configurations $1,2,3$ increase in terms of ``aggressiveness'' from driving slowly with a large headway to fast with a small headway -- with two options $a$ and $b$ for the maximal acceleration.

\begin{remark}
The specific choices are inspired by the following considerations: The implemented road speed limit in Wildau is \unit[50]{km/h} which is approximately \unit[13.9]{m/s}. Vehicles in the original Wildau scenario have a maximal acceleration of \unit[0.8]{m/s$^2$}, while SUMO's default value is \unit[2.6]{m/s$^2$}. Similarly, SUMO's default time headway is \unit[1.0]{s}.
\end{remark}

\textbf{Accident Occurrence.} 
The best estimate for the total number of accidents in Wildau is $\bar{c}_\mathrm{year} \approx 407$. From this, we derive a uniform and a non-uniform accident occurrence model. In both cases, we specify accident probabilities $p^{1,k}$ and $p^{2,k}$ for the binomial model as well as accident intensities $\lambda^{1,k}$ and $\lambda^{2,k}$ for the Poisson model.

\begin{enumerate}
\item \emph{Uniform Accident Occurrence.} Assuming that accidents occur uniformly over the year, we obtain a probability per time bucket of an accident in the system of $p^1 =\bar{c}_\mathrm{year}/N \approx 7.7 \cdot 10^{-4}$. This is the accident probability that we allocate to each traffic scenario $k$. We also suppose that accidents occur uniformly across all vehicles in the system. This implies that the probability that any accident occurs in scenario $k$ within the fleet $\Phi$ is
$$  p^{\Phi, 1,k}  \; = \;    \rho^\Phi\cdot p^1 , \quad k\in\{1,\dots,K\}. $$
This probability is used in the Binomial model. For the Poisson model we set $\lambda^{\Phi,1,k}=p^{\Phi, 1,k}$, $k=1,2, \dots, K$, since the intensity approximately equals the probability of an accident per time bucket. 

In the case of uniform accident occurrence, we do not consider any spatial variations of the likelihood of accidents due to different traffic conditions. This means that we do not distinguish any modules, i.e., we set $R=1$.
 
\item \emph{Non-Uniform Accident Occurrence.} In reality, the likelihood of accidents depends on external factors such as weather and local traffic conditions, e.g., the velocity of vehicles and traffic density. The quantities vary spatially and over time.

From SUMO runs, we obtain for each traffic scenario $k=1,2, \dots, K$ and each module $r=1,2, \dots, R$ pairs $(d_r^k,\bar{v}_r^k)$ on the average density and velocity. These statistics can be computed in SUMO, for example, from data that are obtained at induction loop detectors which are placed within the modules; as a proxy for density, we extract the \emph{occupancy} of the loop detector, i.e., the fraction of time which it is occupied by a vehicle.\\
For $r=1,\dots,R$, we choose benchmark values $d^*_r$ and $\bar{v}^*_r$ for the density and velocity and specify occurrence probabilities and intensities that vary spatially and over time: 
$$
 \lambda^{\Phi, 2,k}_r  \;  =  \; p^{\Phi, 2,k}_r   \; := \; \frac{p^{\Phi, 1,k}}{R} \cdot \frac{\bar{v}_r^k}{\bar{v}_r^*}\cdot\frac{d_r^k}{d_r^*}\cdot e^{-(\zeta^\Phi-1)}  ,\quad k\in\{1,\dots,K\},\quad r\in\{1,\dots,R\}.
$$
The last term refers to deviations of the time headway from SUMO's default value of $\unit[1.0]{s}$: a larger time headway is associated with less risky driving. We set $p^{\Phi, 2,k}=\sum_{r=1}^R p^{\Phi, 2,k}_r$ and  $\lambda^{\Phi, 2,k}=\sum_{r=1}^R \lambda^{\Phi, 2,k}_r$.

In our case studies, we will consider a grid of $R=4$ modules and compute $d_r^k$ and $\bar{v}_r^k$ as averages over measurements from $10$ induction loop detectors that are placed in each module (see also Figure \ref{fig:SUMOWildau2}). We use the scenario averages $d^*_r=\frac{1}{K}\sum_{k=1}^K d_r^k$ and $\bar{v}_r^*=\frac{1}{K}\sum_{k=1}^K \bar{v}_r^k$.
 If $\sum_{k=1}^K \mathbb{E}(\mu^k)  \bar{v}_r^k = \bar{v}_r^* $, $\sum_{k=1}^K \mathbb{E}(\mu^k) d_r^k = d_r^*$, and $\zeta^\Phi =1$, then we essentially recover  on average the case of uniform accident occurrence.
 
 \begin{figure}
 \centering
 \includegraphics[scale=0.15, trim=250 0 450 0, clip]{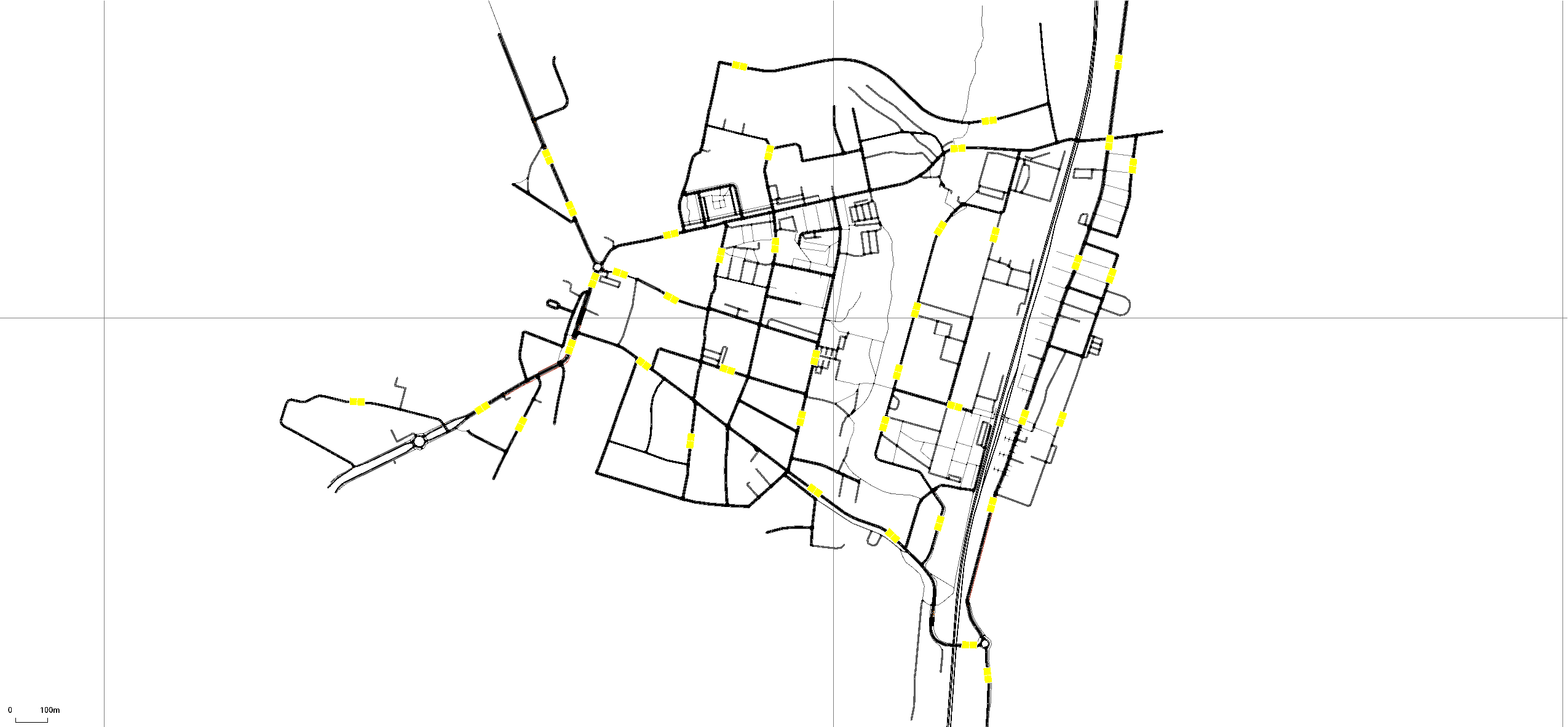}
 \caption{Partition of Wildau and Placement of Induction Loop Detectors.}
 \label{fig:SUMOWildau2}
 
 \end{figure}
 \end{enumerate}

\textbf{Accident Losses.} 
The distributions $F^k$ of accident losses associated with a traffic scenario $k = 1,2, \dots, K$ are constructed on the basis of traffic data that are extracted from the SUMO runs. The general procedure was described in Section~\ref{mtmwal}; here, we explain the specific implementation that we use in our case studies.

The likelihood of accident occurrence was discussed in the previous section. $F^k$ is the conditional distribution of an accident loss in traffic scenario $k$ if an accident occurs. Time in traffic scenario $k$ is enumerated by $t\in [0,T]$, and we assume that the time $\tau$ of the accident conditional on its occurrence is uniformly distributed on $[0,T]$, i.e., 
$\tau\; \sim \; \mathrm{Unif}[0,T].$
We choose a module $\mathcal{R}$ at random in which the accident occurs and assume, respectively, that 
$$P(\mathcal{R}=r) \; = \; \frac{p^{\Phi, \cdot, k}_r}{p^{\Phi, \cdot, k}}, \quad\quad P(\mathcal{R}=r)=\frac{\lambda^{\Phi, \cdot, k}_r}{\lambda^{\Phi, \cdot, k}}, \quad \quad\quad r\in\{1,\dots,R\} . $$
These ratios depend in the case of non-uniform accident occurrence on the specific fleet, since the properties of the fleet alter the route file that is used to generate the SUMO  scenarios; this is true,  although the multiplicative terms $\rho^\Phi$ appear in both the numerator and denominator and cancel out.

In the chosen module we pick one or more vehicles at random, and extract from the traffic scenario data for these vehicles. In our concrete implementation, we simply choose at time $\tau$ a single vehicle $I$ uniformly at random in module $\mathcal{R}$, i.e., its conditional distribution is
$$I\mid \tau,\mathcal{R} \; \sim \; \mathrm{Unif}(\mathcal{M}_\mathcal{R}^\Phi(\tau)). $$
For the purpose of illustrating our approach, the only data we extract are the velocities $v^I$ of the randomly chosen vehicles that are involved in accidents. We set $\psi= v^I$ and assume that the conditional loss distribution $F^{k,\psi} $ is known (we assume that $F^{k,0}$ corresponds to a Dirac measure in $0$; if $\mathcal{M}_\mathcal{R}^\Phi(\tau)=\emptyset$, we set $\psi=0$, resulting in $0$ losses). If we denote the distribution of $\psi$ by $\lcal^{k}$ we obtain the distribution $F^k$ as a mixture
$$ F^k \; = \; \int F^{k, \psi} \;d  \lcal^{k}  .$$

In our case studies, we will assume that $F^{k, \psi} = F^\psi$ for all $k$; however, the mixing distribution $\lcal^{k}$ will dependend on the traffic scenario $k$. We consider the following examples for $F^\psi$:

\begin{enumerate}
\item \emph{Gamma Distribution.} We define distributions with varying levels of \emph{dispersion}. A measure for the dispersion of a random variable $X$ is the coefficient of variation defined by $c_v=\sqrt{\mathrm{Var}(X)}/\mathbb{E}(X)$. For $c_v\in\{1/2,~1,~2\}$, we choose $F^\psi  \; =  \;   \Gamma \left( \frac{1}{c_v^2} , \; \frac{1}{c_v^2 \psi^2} \right).$ The expectation of this distribution is $\psi^2$ and increases quadratically with $\psi$, the velocity of the vehicle involved in an accident; this is consistent with the fact that losses scale with kinetic energy. The variance of the distribution $F^\psi$ equals $c_v^2\psi^4$, hence the coefficient of variation is indeed $c_v$.
\item \emph{Log-Normal Distribution.} We consider log-normal distributions with expectation $\psi^2$ and variance $c_v^2 \psi^4$, implying that the coefficient of variation is again $c_v\in\{1/2,~1,~2\}$. This log-normal distribution is obtained as the distribution of $\exp(Z)$ for a normal random variable $Z$ with expectation $\ln(\psi^2/\sqrt{1+c_v^2})$ and variance $\ln(1+c_v^2)$, i.e.,
$$F^\psi  \; =  \;  \mathcal{LN}\left(  \ln \left( \frac{\psi^2}{\sqrt{1+c_v^2}}   \right)   , \; \ln(1+c_v^2) \right)  .$$
\end{enumerate}

\begin{remark}
The conditional frequency and severity given traffic scenarios is an essential ingredient to our modeling approach. Its flexibility comes from the fact that traffic can be micro-simulated, and its impact on aggregate losses can be analyzed if conditional loss distributions are available. These need to be studied in more detail on the basis of empirical data and structural considerations. This paper focuses on an illustration of the simulation methodology. Reviews on different methodologies for the assessment of accident frequencies and severities based on underlying covariates are provided by \textcite{LORD2010291}, \textcite{SAVOLAINEN20111666}, \textcite{MANNERING20141}, and \textcite{THEOFILATOS2014244}. \textcite{LIAN2020105711} reviews the use of big data for the analysis of traffic conditions and their relationships to accident frequencies and severities. \textcite{Retallack:2019aa} review articles on the relationship of traffic congestion and accidents. \textcite{MALIN2019181} investigate the relative accident risk of different road and weather conditions and combinations of conditions using data for major roads in Finland; their analysis is based on the notion of Palm probability. Using generalized additive models, \textcite{Becker:2022aa}  quantify the combined effects of traffic volume and meteorological parameters on probabilities of 78 different crash types. \textcite{COMI2022798} investigate the suitability of various data mining techniques in analyzing the factors underlying accidents and predicting these in case studies based on data collected in Rome. 
\end{remark}

\subsection{Case Studies}

\subsubsection{Overview}

We illustrate our modeling approach in case studies on multiple levels. A selection of case studies is discussed in detail in Sections \ref{eng_per} \& \ref{sec:actuarial}. All numerical results for the following choices are documented online in tables in Appendix~\ref{sec:tables}:
\begin{enumerate}
\item \emph{Fleet Models.} We analyze six driving configurations with three different fleet proportions.
\item \emph{Accident Occurrence.} In our traffic system accidents occur uniformly or non-uniformly in space. Their number is given by a Binomial or a Poisson model with parameters depending on traffic condition.
\item \emph{Accident Losses.} We study two parametric families of loss distributions with three different choices for the coefficient of variation.
\item \emph{Insurance Design.} Insurance losses are a function of the total losses; we distinguish three contract designs.
\end{enumerate}
Denoting aggregate losses by $L$, we evaluate for each type of insurance coverage $h$ the resulting insurance losses $h(L)$ in terms of their expectation, variance, and skewness, and the monetary risk measures Value-at-Risk and Average Value-at-Risk, also called Expected Shortfall. To analyze the distributions in detail, we provide qq-plots and estimates of cumulative distribution functions and densities. The main tool to access the random variable $h(L)$ is Monte Carlo sampling; we provide a pseudo-code how we obtain samples of $L$ in Algorithm \ref{alg:lossSampling} in Appendix \ref{sec:sampling}. In Section~\ref{sec:actuarial}, we compare this approach to the Normal mean-variance-mixture approximation introduced in Section~\ref{sec:approximation}. 

This paper explores the analysis and management of risks that occur in vehicle fleets in traffic systems. We distinguish to perspectives:
\begin{enumerate}
\item \emph{The Engineering Perspective.} In these case studies, we fix the accident occurrence and  accident loss distributions and vary the fleet models, i.e., driving configurations and fleet proportions. We focus on $\mathbb{E}(L)$, $\mathrm{Var}(L)$, and complement these with analyses of the performance of traffic system.
\item \emph{The Actuarial Perspective.} In these case studies, we fix the fleet model and vary accident occurrence and accident loss distributions as well as the insurance design. We study the distribution of $L$ and the insurance prices $\mathbb{E}(h(L))$. 
\end{enumerate}

\subsubsection{The Engineering Perspective}\label{eng_per}

Our micro-modeling approach allows us to study the effects of different traffic-related controls on total losses $L$; we investigate the effects of fleet size and traffic configuration. Throughout this section, we consider non-uniform accident occurrence in the Binomial model with Gamma distributed accident losses and a coefficient of variation $c_v=1$. 

\begin{figure}

\begin{minipage}{0.5\textwidth}
\centering
\includegraphics[scale=0.8]{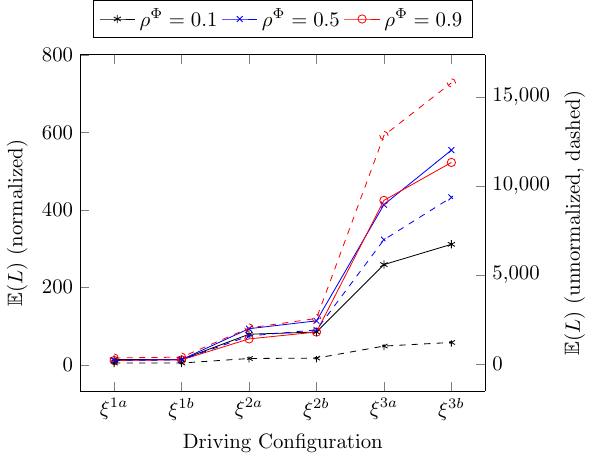}
\subcaption{Expectation of the Total Loss}
\label{fig:engineering1a}
\end{minipage}\begin{minipage}{0.5\textwidth}
\centering
\includegraphics[scale=0.8]{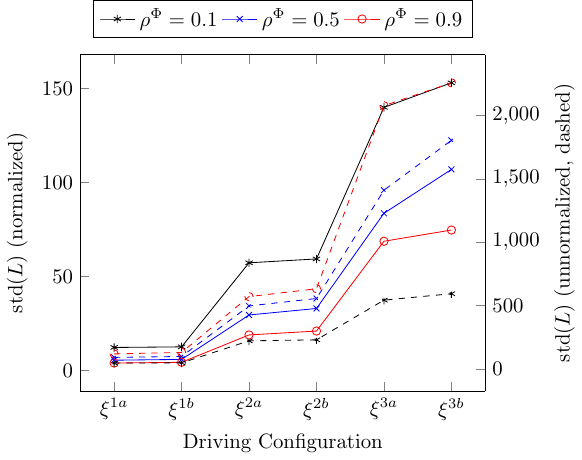}
\subcaption{Standard Deviation of the Total Loss}
\label{fig:engineering1b}
\end{minipage}

\caption{Impact of Fleet Size and Driving Configuration on the Total Loss $L$ (solid lines represent normalized values (left y-axis), dashed lines unnormalized ones (right y-axis)).}
\label{fig:engineering1}
\end{figure}

\textbf{Losses.} We evaluate expected loss $\mathbb{E}(L)$ and standard deviation $\mathrm{std}(L)$ for different fleet models. To compare losses for different fleet sizes, we normalize losses per $100$ expected insured vehicles: For each traffic scenario $k$, the number of insured vehicles is the number of vehicles belonging to $\Phi$ as given in the underlying route file. We refer to Appendix \ref{sec:tables} for more details. The model specification was explained in Section~\ref{sec:speci} which includes in particular a description of the driving configurations. The results are documented in Figure \ref{fig:engineering1}. The solid lines are the normalized quantities.

In Figure \ref{fig:engineering1a}, we see that increasing the aggressiveness of driving increases both the total and normalized expected loss. An impact of the maximal acceleration on losses is only substantial for the most aggressive driving configurations $\xi^{3\cdot}$. Increasing the fleet size increases the expected loss which is primarily due to the fact that we count losses only within the fleet and a higher volume is associated with higher losses. More interesting is the normalized case: apparently higher speeds also increase the normalized losses.

In Figure \ref{fig:engineering1b}, we have the corresponding standard deviations. Increasing the aggressiveness of driving increases the standard deviation of the total loss. The standard deviations of the normalized losses are decreasing in the fleet size. The main reason is that fluctuations normalized for a fixed volume are larger for smaller pools than for larger pools; a rational for this is provided by the law of large numbers and the central limit theorem.

\begin{figure}

\begin{minipage}{0.5\textwidth}
\centering
\includegraphics[scale=0.8]{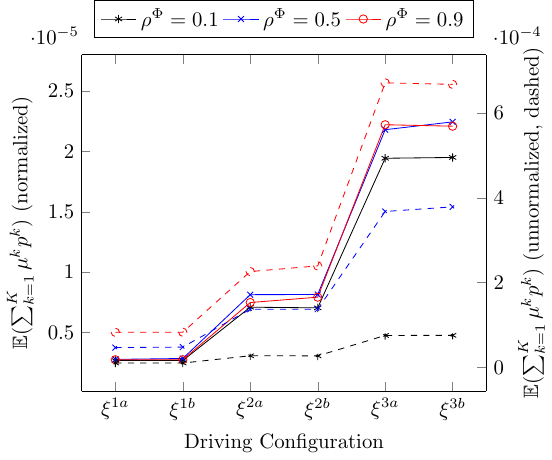}
\subcaption{Expectation}
\label{fig:engineeringFa}
\end{minipage}\begin{minipage}{0.5\textwidth}
\centering
\includegraphics[scale=0.8]{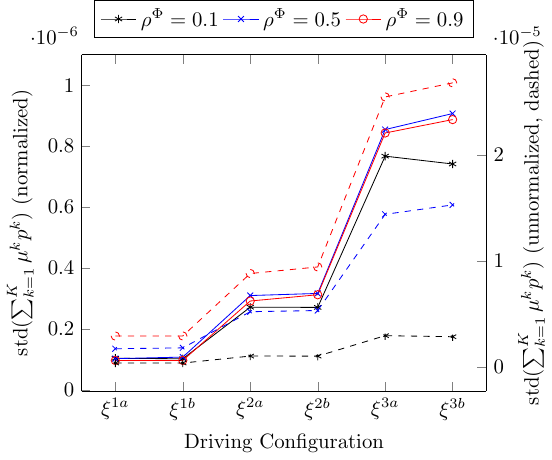}
\subcaption{Standard Deviation}
\label{fig:engineeringFb}
\end{minipage}

\caption{Impact of Fleet Size and Driving Configuration on Expectation and Standard Deviation of Average Accident Frequency (solid lines represent normalized values (left y-axis), dashed lines unnormalized ones (right y-axis)).}
\label{fig:engineeringF}
\end{figure}

\begin{figure}

\begin{minipage}{0.5\textwidth}
\centering
\includegraphics[scale=0.8]{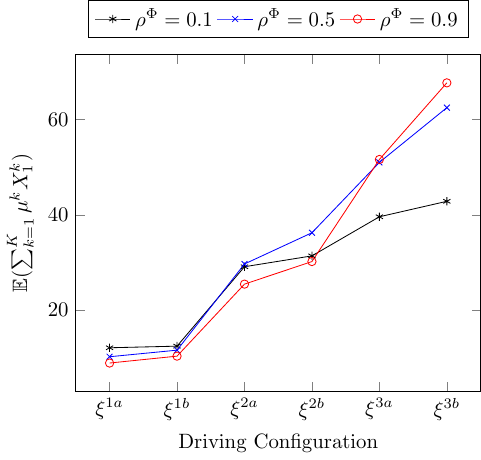}
\subcaption{Expectation}
\label{fig:engineeringSa}
\end{minipage}\begin{minipage}{0.5\textwidth}
\centering
\includegraphics[scale=0.8]{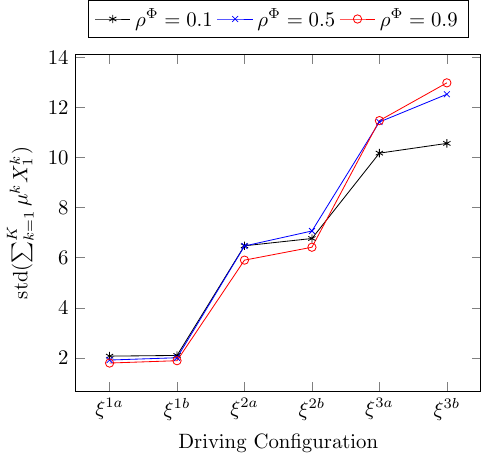}
\subcaption{Standard Deviation}
\label{fig:engineeringSb}
\end{minipage}

\caption{Impact of Fleet Size and Driving Configuration on Expectation and Standard Deviation of Average Accident Severity.}
\label{fig:engineeringS}
\end{figure}

The frequency and severity of accidents are, of course, increasing in the aggressiveness of driving. To demonstrate this, we evaluate the expectation and the standard deviation of the average accident frequency $\sum_{k=1}^K\mu^kp^k$ (both normalized and unnormalized) and the average accident severity $\sum_{k=1}^K \mu^k X_1^k$, as displayed in Figure \ref{fig:engineeringF} and Figure \ref{fig:engineeringS}. A larger fleet increases frequency and, in aggressive scenarios, also the expected accident severity. This is, of course, due to the specific choice of the driving behavior of the considered fleets in comparison to the driving behavior of the remaining vehicles and does not necessarily hold for all traffic systems in general. In this section, we discussed losses for selected special cases. A comprehensive set of tables for all other cases and statistical functionals is provided online in Appendix \ref{sec:tables}.

Changing the characteristics of the fleet not only affects the losses. At the same time, this has an impact on the performance of the traffic system. We present further results in Appendix \ref{sec:systemPerformance}.

\subsubsection{The Actuarial Perspective}\label{sec:actuarial}

From an actuarial perspective, it is relevant to understand the \emph{risk} that corresponds to the insurance losses. This requires a more detailed analysis of the \emph{probability distributions}. To do this, we pick a particular fleet model and use probabilistic techniques to evaluate the distribution of $h(L)$. From now on, we consider $\rho^\Phi=0.5$ with driving configuration $\xi^{2a}$ and non-uniform accident occurrence.

\textbf{Distributional Analysis of Losses.} We start our investigations with the total losses $L$. Table \ref{table:functionals} shows the evaluation of statistical functionals for different accident losses in the case of the Binomial model. These numbers quantify the risk entailed in the total losses. Both the distributional family and the chosen coefficient of variation for the accident loss model have a substantial effect on the risk.

\begin{table} 
\centering 
\scalebox{0.8}{ 
\begin{threeparttable} 
\caption[]{Statistical Functionals of $L$ for $\rho^\Phi=0.5$ and $\zeta^{2a}$.} 
\label{table:functionals}
\begin{tabular}{rrrrrrrr} \toprule 
& \multicolumn{7}{c}{Binomial Model}    \\ \midrule 
& \multicolumn{3}{c}{Gamma}  && \multicolumn{3}{c}{Log-Normal}   \\ \midrule 
& $c_v=0.5$ & $c_v=1.0$ & $c_v=2.0$  && $c_v=0.5$ & $c_v=1.0$ & $c_v=2.0$   \\     \toprule 
 \\ 
$\mathbb{E}(L)$ &    1577.8 & 1571.5 & 1578.4 && 1581.7 & 1576.2 & 1582.7\\ 
$\mathrm{Var}(L)$&    160179.5 & 247943.9 & 626190.6 && 162067.5 & 247069.8 & 614713.4 \\ 
$\varsigma_L$ &   0.333 & 0.561 & 0.993 && 0.377 & 0.660 & 2.152  \\ 
$\mathrm{VaR}_{0.9}(L)$  &    2111.9 & 2219.1 & 2634.2 && 2112.9 & 2242.4 & 2526.6  \\ 
$\mathrm{ES}_{0.9}(L)$ &    2331.7 & 2538.8 & 3233.3 && 2342.4 & 2557.5 & 3280.3  \\ 
$\mathrm{VaR}_{0.95}(L)$ &    2275.9 & 2471.3 & 3070.5 && 2288.4 & 2468.1 & 2996.2  \\ 
$\mathrm{ES}_{0.95}(L)$ &    2468.9 & 2748.0 & 3628.9 && 2491.4 & 2772.5 & 3838.4  \\ 
$\mathrm{VaR}_{0.99}(L)$ &    2587.3 & 2891.3 & 3999.6 && 2608.8 & 2943.8 & 4228.6  \\ 
$\mathrm{ES}_{0.99}(L)$ &    2755.9 & 3188.6 & 4490.2 && 2809.0 & 3234.4 & 5424.2 \\ 
\bottomrule 
\end{tabular} 
\begin{tablenotes} 
\small 
\item The statistical functionals of the total loss are approximated using $10,000$ independent samples of $L$.
\end{tablenotes} 
\end{threeparttable} 
}
\label{tab1}

\end{table}

\begin{figure}
\begin{minipage}{0.3\textwidth}
\centering
\includegraphics[scale=0.6]{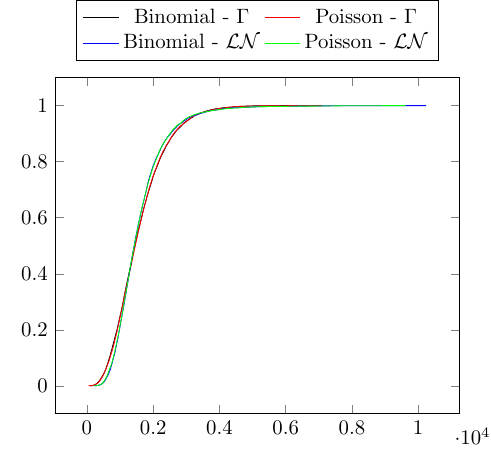}
\subcaption{CDF}
\label{fig:actuarial1edf}
\end{minipage}\hspace{0.5cm}\begin{minipage}{0.3\textwidth}
\centering
\includegraphics[scale=0.6]{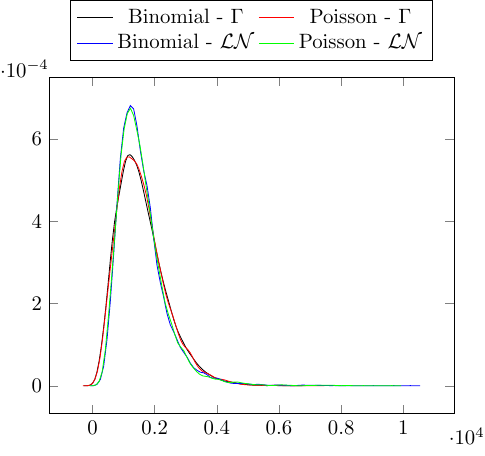}
\subcaption{Density}
\label{fig:actuarial1density}
\end{minipage}\hspace{0.5cm}\begin{minipage}{0.3\textwidth}
\centering
\includegraphics[scale=0.6]{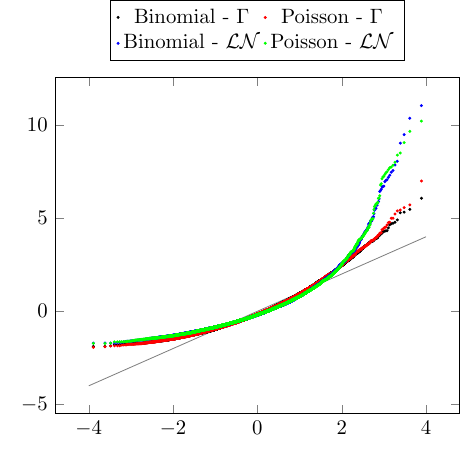}
\subcaption{QQ-Plot}
\label{fig:actuarial1qq}
\end{minipage}
\caption{Distribution of the Total Loss for Fixed Coefficient of Variation $c_v=2$.}
\label{fig:actuarial1}
\end{figure}

A  visual impression of the distributions is provided in Figure \ref{fig:actuarial1}. We compare different accident loss models while fixing the coefficient of variation $c_v=2$. We plot the empirical distribution functions as estimates of the cumulative distribution function (CDF) and a kernel density estimate of the corresponding densities. Moreover, Figure \ref{fig:actuarial1qq} shows qq-plots for the quantiles of standardized values of the losses against quantiles of a standard Normal distribution (samples are standardized by subtracting their sample average and dividing by their sample standard deviation). 

We find that the Binomial and the Poisson model do not differ too much. Yet, log-normal accident losses produce heavier tails than the corresponding Gamma losses. The qq-plots reveal that the right tails are heavier compared to a Normal distribution while the left tails are lighter. The latter observation simply relates to the fact that the original losses are non-negative while the Normal distribution takes values on the whole real line.

\begin{figure}
\begin{minipage}{0.3\textwidth}
\centering
\includegraphics[scale=0.6]{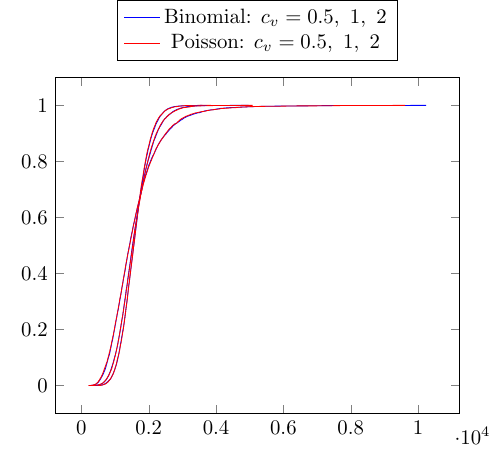}
\subcaption{CDF}
\label{fig:actuarial2edf}
\end{minipage}\hspace{0.5cm}\begin{minipage}{0.3\textwidth}
\centering
\includegraphics[scale=0.6]{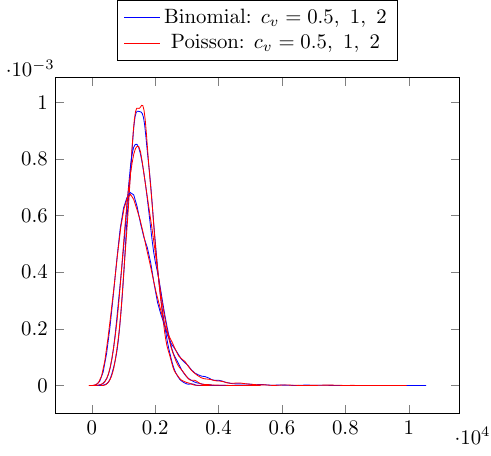}
\subcaption{Density}
\label{fig:actuarial2density}
\end{minipage}\hspace{0.5cm}\begin{minipage}{0.3\textwidth}
\centering
\includegraphics[scale=0.6]{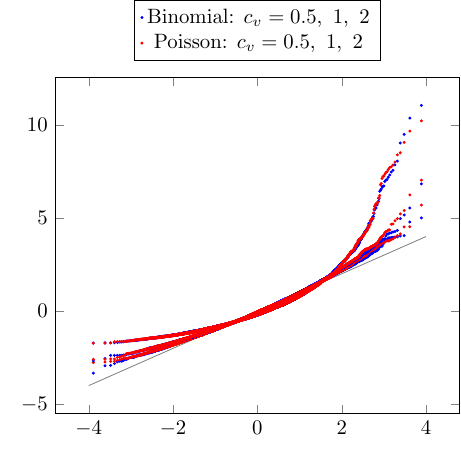}
\subcaption{QQ-Plot}
\label{fig:actuarial2qq}
\end{minipage}
\caption{Distribution of the Total Loss for Log-Normal Accident Losses and Varying Coefficient of Variation.}
\label{fig:actuarial2}
\end{figure}

In Figure \ref{fig:actuarial2}, we analyze the impact of the coefficient of variation while fixing the log-normal distribution for the accident losses. We see again that Binomial and Poisson model do not differ substantially. However, the effect of the coefficient of variation is clearly visible: increasing $c_v$ produces heavier right tails. Introducing dispersion to accident losses substantially changes the distribution of the total losses.

\textbf{Comparison of Losses and Normal Mean-Variance-Mixture Approximation.} In Section \ref{sec:approximation}, we suggested a Normal mean-variance-mixture approximation for the total loss. To study the quality of this approximation, we generate $10,000$ samples from the approximation; in the following, we focus on the case of Gamma distributed accident losses with coefficient of variation $c_v=1$.

To sample from the approximation, we rely on the following computations of $m^k$ and $(\sigma^k)^2$. Using $\mathbb{E}(\mathbf{1}_c^k)=p^k$ and $\mathbb{E}(X_c^k) = \mathbb{E}(\mathbb{E}(X_c^k\mid \psi)) = \mathbb{E}(\psi^2) = \int \psi^2 d \lcal^k $, we obtain for the Gamma losses $m^k =p^k\cdot \int \psi^2 d \lcal^k $ and $(\sigma^k)^2 = p^k\cdot c_v^4\cdot \int \psi^4 d \lcal^k \cdot  \left(1+\frac{1}{c_v^2}\right)\frac{1}{c_v^2}-(m^k)^2$. 
\begin{remark}
For the notation, we refer to Sections \ref{sec:approximation} \& \ref{sec:speci}. The computations are valid for any coefficient of variation $c_v$, but we use only $c_v=1$ in the numerical case study.
\end{remark}
The involved moments of $\psi$ are approximated using $10,000$ samples from the traffic simulation, for each $k=1,\dots,K$. A sample from the Normal mean-variance-mixture approximation is generated by, first, sampling $\mu$ and, second (conditional on $\mu$), sampling the normal random variable $\sum_{k=1}^K N\mu^k m^k + Z $ with $Z  \sim  \ncal \left(  0 , \; \sum_{k=1}^K  N\mu^k {\left(\sigma^k\right)}^2   \right).$

\begin{figure}
\centering
\includegraphics[scale=0.8]{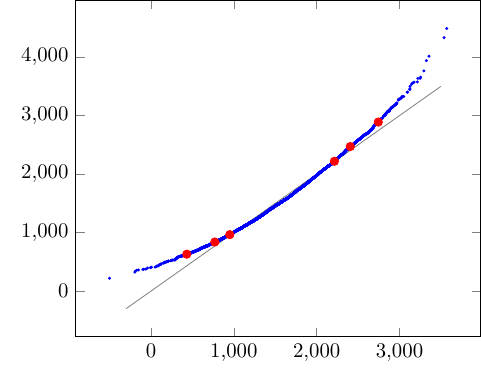}
\caption{QQ-Plot of $10,000$ Monte Carlo Samples (y-axis) vs $10,000$ Samples of Mixture Approximation (x-axis) (the red dots mark the $1\%,~5\%,~10\%,~90\%,~95\%,~99\%$ quantiles).}
\label{fig:actuarial3}
\end{figure}

Figure \ref{fig:actuarial3} shows the qq-plot comparing quantiles of the crude Monte Carlo simulation with quantiles of the approximation. This demonstrates the quality of our suggested approximation. It is almost exact between the $5$\,\% and $95$\,\% quantile as the values lie on the halfline. It is still very good for the $1-5$\,\% and $95-99$\,\% quantiles and is only less accurate in the extreme tails where also in the Monte Carlo simulation only few data points are available. These analyses, on the one hand, confirm the postulated structural model insight. On the other hand, they also validate the implementation of our crude Monte Carlo sampling.

\textbf{Pricing and Evaluation Methods.} To conclude our case studies, we study prices for various insurance contracts. We compare $\mathbb{E}(L)$ (full coverage), $\mathbb{E}(\max(L-\theta,0))$ (constant deductible), and $\mathbb{E}(\min(L,\theta))$ (stop loss) for different values of $\theta$. 
\begin{figure}

\centering
\includegraphics[scale=0.8]{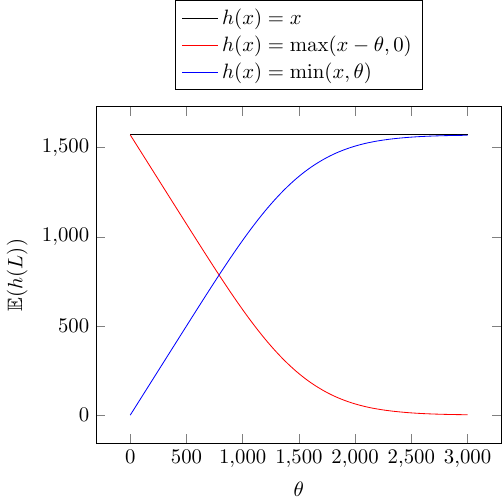}
\caption{Insurance Prices.}
\label{fig:actuarial4}
\end{figure}
The results are given in Figure \ref{fig:actuarial4}. We obtain the typical hockey stick profiles satisfying the parity $\mathbb{E}(L)=\mathbb{E}(\max(L-\theta,0))+\mathbb{E}(\min(L,\theta))$. We note that other insurance contracts can easily be represented in our framework; also deductibles per accident can be implemented by changing the accident loss distributions accordingly.

Besides Monte Carlo methods, our Normal mean-variance mixture approximation and the correction suggested in Section \ref{sec:approximation} provide alternative techniques to compute the prices $\mathbb{E}(h(L))$ for the different types of coverage $h$. To compute the correction term $C_h$, we need to compute $(\zeta^k)^3$ (for the notation we refer to Sections \ref{sec:approximation} \& \ref{sec:speci}). For Gamma distributed losses, we obtain
\begin{equation*}
(\zeta^k)^3=p^k\cdot c_v^6\cdot \int \psi^6 d \lcal^k \cdot \left(2+\frac{1}{c_v^2}\right)\left(1+\frac{1}{c_v^2}\right)\frac{1}{c_v^2} - 3m^k(\sigma^k)^2-(m^k)^3.
\end{equation*}
Since $\mathbb{E}(h(L))=\mathbb{E}(\mathbb{E}(h(L)\mid\mu))$, we may generate samples of $\mu$ and evaluate the corresponding conditional expectations $\mathbb{E}(h(L)\mid\mu)$; in the Normal mean-variance mixture approximation, these are expectations of functions of normally distributed random variables. We compute these expectations numerically as integrals with respect to Lebesgue measure using a normal density.

\begin{figure}
\begin{minipage}{0.5\textwidth}
\centering
\includegraphics[scale=0.7]{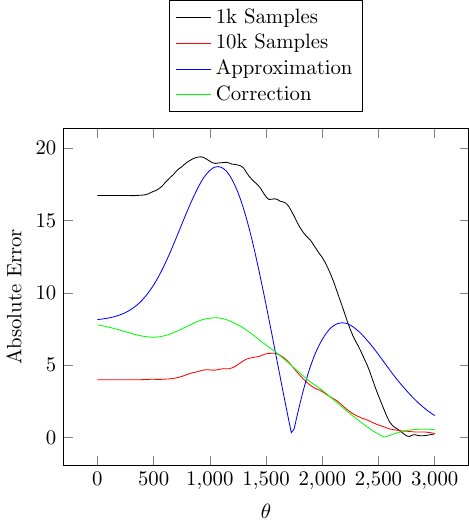}
\subcaption{Absolute Error}

\end{minipage}\hspace{0.5cm}\begin{minipage}{0.5\textwidth}
\centering
\includegraphics[scale=0.7]{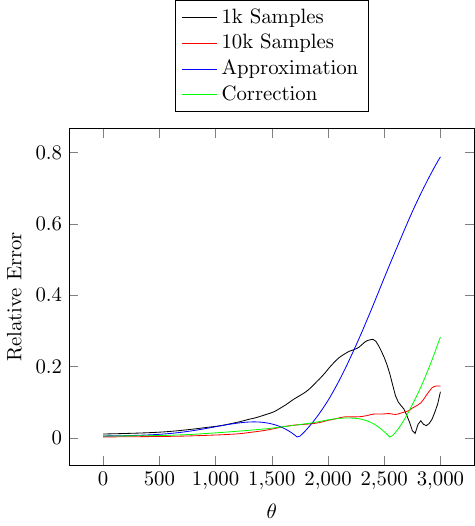}
\subcaption{Relative Error}
\end{minipage}

\caption{Comparison of Estimation Errors.}
\label{fig:actuarial5}
\end{figure}

We compare the estimation errors of the different approaches in Figure \ref{fig:actuarial5}. We produce $100,000$ samples of $L$ to approximate the ``true'' value of $\mathbb{E}(\max(L-\theta,0))$ of coverage with constant deductible, for different values of $\theta$. Independently, we generate $10,000$ samples and consider Monte Carlo approximations based on all $10,000$ samples, and based only on the first $1,000$ samples. We also study the Normal mean-variance mixture approximation with and without correction using $1,000$ samples of $\mu$.

While the absolute error generally decreases in the deductible $\theta$ for all methods (apart from some local effects), the relative error increases for larger values of $\theta$ which are associated with a lower price of the contract. At the same time, we observe that, in terms of the relative error, the Normal mean-variance mixture approximation produces reasonable estimation results (compared to $1,000$ samples) for moderate values of $\theta$. This is in line with our previous observations on the quality of our Normal mean-variance mixture approximation which becomes worse in the extreme tail of $L$. Interestingly, the estimation error can largely be reduced using the correction; with the correction, the estimation becomes quite good even for large values of $\theta$.

\section{Conclusion}\label{sec:conclude}

This paper developed a methodology to study accident losses based on microscopic traffic simulators. An adaption of the digital twin paradigm enabled us to test the impact of fleet sizes and their driving configuration on system efficiency, accident losses, and insurance premiums. It was shown that -- on a one-year horizon -- total losses can be approximated by a mean-variance mixture of normal distributions. This offered an alternative technique to evaluate the model; the numerical efficiency can be increased adding a correction term that is derived by Stein's method. The proposed methodology can be extended and modified, for example, and utilized to study future traffic systems. We illustrated in counterfactual case studies that were based on the software SUMO how accident risk can be successfully analyzed, both from an engineering and an actuarial perspective. 

Future research should also address the important issues of calibration and validation. While real data can be used to calibrate models that describe historical and current transportation systems,  simulation models that generate artificial data are essential to evaluate new technologies in future transportation systems. An important question is to what extent and how historical data can be methodically used to calibrate and validate such simulation models. For example, real microscopic traffic data, e.g., on accident patterns, collected by means of telematics technologies, could be applied to optimize models in the future. The comparison of simulated and real data will also allow to investigate the impact of traffic conditions on accident occurrence, financial losses, and the probability distributions thereof.

\printbibliography

\vspace{1cm}

\textsc{Sojung Kim, Institute of Actuarial and Financial Mathematics \& House of Insurance, Leibniz Universität Hannover, Welfengarten 1, 30167 Hannover, Germany,} \emph{email:} \href{mailto:sojung.kim@insurance.uni-hannover.de}{sojung.kim@insurance.uni-hannover.de}\\

\textsc{Marcel Kleiber, Institute of Actuarial and Financial Mathematics \& House of Insurance, Leibniz Universität Hannover, Welfengarten 1, 30167 Hannover, Germany,} \emph{email:}  \href{mailto:marcel.kleiber@insurance.uni-hannover.de}{marcel.kleiber@insurance.uni-hannover.de}\\

\textsc{Stefan Weber, Institute of Actuarial and Financial Mathematics \& House of Insurance, Leibniz Universität Hannover, Welfengarten 1, 30167 Hannover, Germany,} \emph{email:} \href{mailto:stefan.weber@insurance.uni-hannover.de}{stefan.weber@insurance.uni-hannover.de}

\pagebreak

\appendix

\section{Supplementary Material: Literature Review}
\label{sec:appendixReview}

In this online supplement, we include further references to the literature on traffic microsimulation. Microscopic traffic models can clearly be used to improve traffic efficiency in complex networks. Leveraging the simulator Aimsun, \textcite{Osorio2013} address questions of this type; the authors develop a stochastic optimization framework based on coupling the Aimsun simulator with a metamodel to optimize signal plans in a city. \textcite{Osorio2015} extend the microscopic traffic model for fuel consumption to determine energy-efficient traffic management strategies. In addition to model building (if data are available), whether macroscopic or microscopic, calibration can be a challenge. \textcite{Floetteroed2011a} propose a Bayesian approach to calibrating travel demand. \textcite{Zhang2017} discuss the calibration of large-scale traffic simulators; \textcite{Osorio2019} focus on the calibration of car-following models for the simulation of a traffic network.

In addition to the traditional focus on efficiency, there is another strand of literature that evaluates the safety of transportation systems. Up to now, mainly historical data have been used to examine accident frequency and severity. For reviews, we refer to \textcite{Lord2010} and \textcite{Tsoi2015}. Statistical modeling approaches permit inference when sufficient data are available on the level of the granularity of the analysis. For example, \textcite{Yu2019} estimate the impact of microscopic traffic variables on crash risks.  \textcite{Ortelli2021} analyze the impact of public traffic policies on crash severity. 

In the absence of  data, physical models of traffic can be used to generate artificial data. In our research group, we have shown how perceptual errors can be added to microscopic traffic models to endogenously model the occurrence of accidents (cf. \textcite{Berkhahn2018} and \textcite{Berkhahn2022}) -- a topic that is particularly relevant for sensors of autonomous vehicles. The models allow characterizing the trade-off between safety and efficiency of transportation systems. 

\section{Supplementary Material: References to SUMO Documentation}
\label{sec:sumoReferences}

We provide specific references to the SUMO documentation regarding the generation of scenarios in Section \ref{sec:SumoScenario}.

\begin{itemize}
\item \textbf{Network}
\begin{itemize}
\item Overview on network generation: \url{sumo.dlr.de/docs/index.html\#network_building}: 
\item Graphical network editor: \url{sumo.dlr.de/docs/Netedit/index.html}
\item Import from OpenStreetMap: \url{sumo.dlr.de/docs/Networks/Import/OpenStreetMap.html}
\end{itemize}
\item \textbf{Demand}
\begin{itemize}
\item Overview on traffic demand modeling: \url{sumo.dlr.de/docs/Demand/Introduction_to_demand_modelling_in_SUMO.html}
\item Generation of traffic demand: \url{sumo.dlr.de/docs/Demand/Activity-based_Demand_Generation.html}
\item Automated generation of a route file: \url{sumo.dlr.de/docs/Demand/activitygen.html}
\end{itemize}
\item \textbf{Scenarios}
\begin{itemize}
\item Vehicles: \url{sumo.dlr.de/docs/Definition_of_Vehicles.html}
\item Publicly available SUMO scenarios: \url{sumo.dlr.de/docs/Data/Scenarios.html}
\end{itemize}
\item \textbf{Randomness}
\begin{itemize}
\item Overview on randomness: \url{sumo.dlr.de/docs/Simulation/Randomness.html}
\end{itemize}
\end{itemize}

\section{Supplementary Material: Traffic System Performance}
\label{sec:systemPerformance}

\begin{figure}
\begin{center}
\begin{minipage}[t]{0.3\textwidth}
\centering
\includegraphics[scale=0.55]{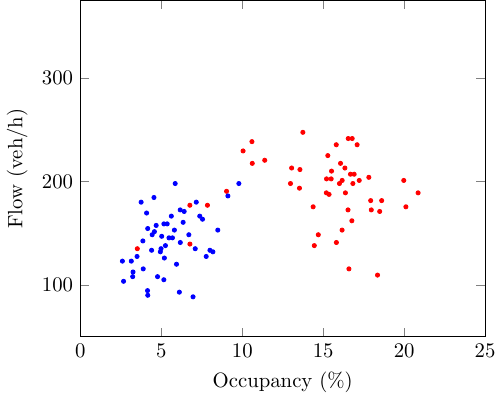}
\subcaption{$\rho^\Phi=0.1$, $\xi^{1b}$}
\end{minipage}\begin{minipage}[t]{0.3\textwidth}
\centering
\includegraphics[scale=0.55]{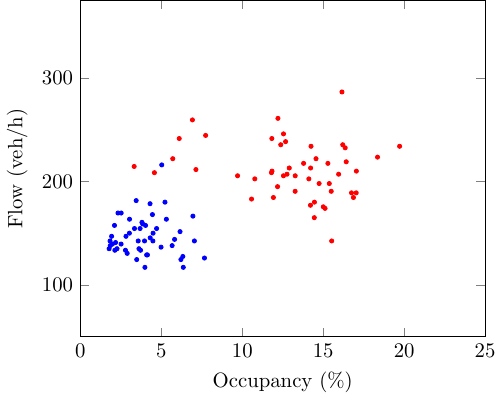}
\subcaption{$\rho^\Phi=0.1$, $\xi^{2b}$}
\end{minipage}\begin{minipage}[t]{0.3\textwidth}
\centering
\includegraphics[scale=0.55]{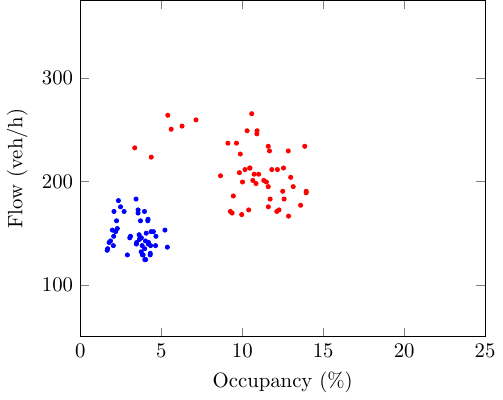}
\subcaption{$\rho^\Phi=0.1$, $\xi^{3b}$}
\end{minipage}
\\[4ex]
\begin{minipage}[t]{0.3\textwidth}
\centering
\includegraphics[scale=0.55]{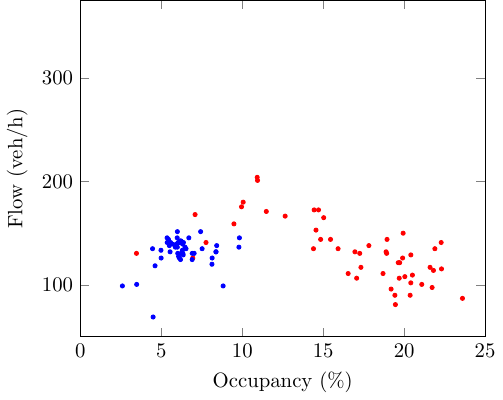}
\subcaption{$\rho^\Phi=0.9$, $\xi^{1b}$}
\end{minipage}\begin{minipage}[t]{0.3\textwidth}
\centering
\includegraphics[scale=0.55]{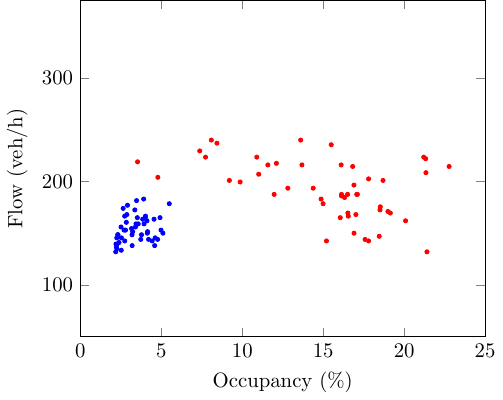}
\subcaption{$\rho^\Phi=0.9$, $\xi^{2b}$}
\end{minipage}\begin{minipage}[t]{0.3\textwidth}
\centering
\includegraphics[scale=0.55]{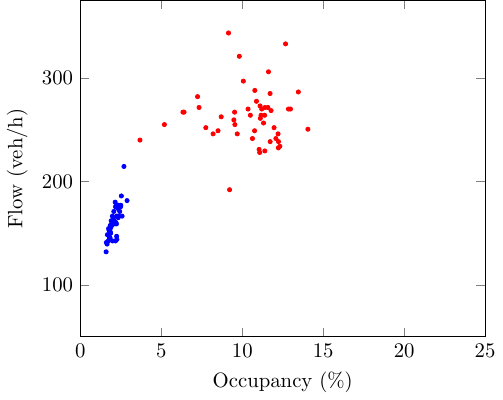}
\subcaption{$\rho^\Phi=0.9$, $\xi^{3b}$}
\end{minipage}
\end{center}
\caption{Flow-Occupancy Fundamental Diagrams (high traffic volume scenarios are highlighted in red and low traffic volume scenarios in blue).}
\label{fig:engineeringFD-flow}
\end{figure}

\begin{figure}
\begin{center}
\begin{minipage}[t]{0.3\textwidth}
\centering
\includegraphics[scale=0.55]{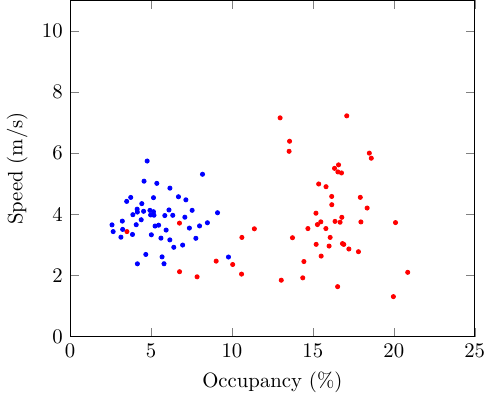}
\subcaption{$\rho^\Phi=0.1$, $\xi^{1b}$}
\end{minipage}\begin{minipage}[t]{0.3\textwidth}
\centering
\includegraphics[scale=0.55]{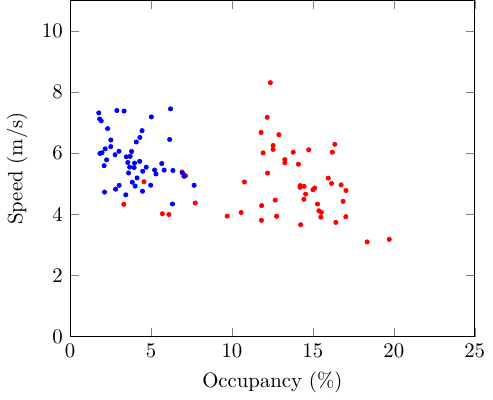}
\subcaption{$\rho^\Phi=0.1$, $\xi^{2b}$}
\end{minipage}\begin{minipage}[t]{0.3\textwidth}
\centering
\includegraphics[scale=0.55]{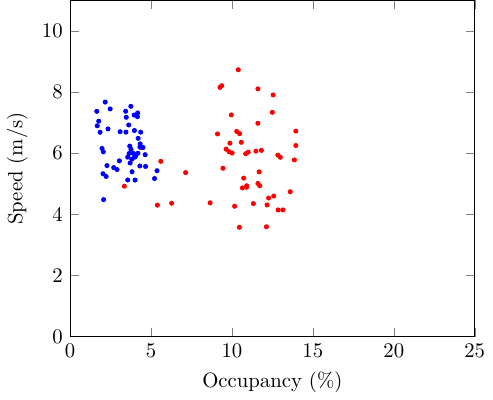}
\subcaption{$\rho^\Phi=0.1$, $\xi^{3b}$}
\end{minipage}
\\[4ex]
\begin{minipage}[t]{0.3\textwidth}
\centering
\includegraphics[scale=0.55]{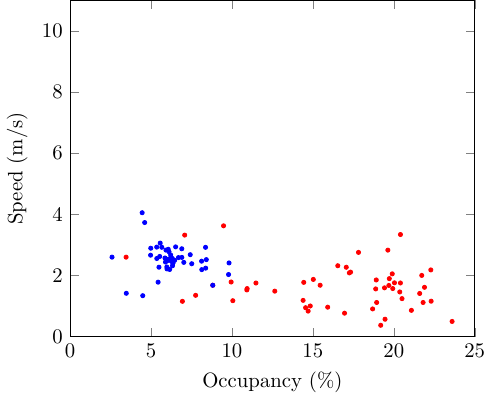}
\subcaption{$\rho^\Phi=0.9$, $\xi^{1b}$}
\end{minipage}\begin{minipage}[t]{0.3\textwidth}
\centering
\includegraphics[scale=0.55]{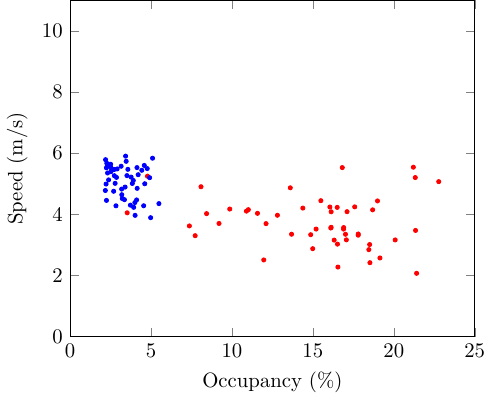}
\subcaption{$\rho^\Phi=0.9$, $\xi^{2b}$}
\end{minipage}\begin{minipage}[t]{0.3\textwidth}
\centering
\includegraphics[scale=0.55]{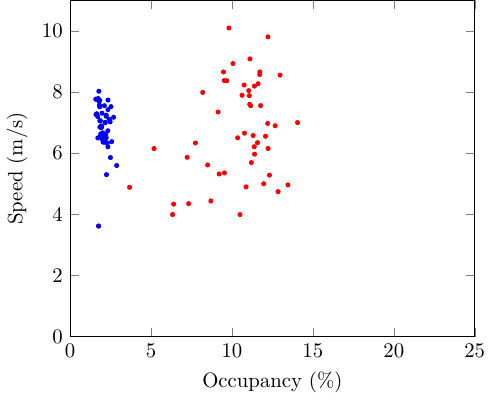}
\subcaption{$\rho^\Phi=0.9$, $\xi^{3b}$}
\end{minipage}
\end{center}
\caption{Speed-Occupancy Fundamental Diagrams (high traffic volume scenarios are highlighted in red and low traffic volume scenarios in blue).}
\label{fig:engineeringFD-speed}
\end{figure}

In this section, we investigate the impact of changes in fleet characteristics on the performance of the traffic system. Using the $40$ induction loop detectors placed in the traffic system, we evaluate the values  of selected traffic statistics (flow, average speed, and occupancy). Denoting them by $\chi_k^1,\dots,\chi_k^{40}$ for each scenario $k=1,\dots,K$, we compute averages $\bar{\chi}_k=1/40\sum_{i=1}^{40} \chi_k^i$. Pairing flow-occupancy values and speed-occupancy values yields empirical fundamental diagrams on the urban level (see, e.g., \textcite{Geroliminis2008}). For the purpose of data exploration, we draw scatter plots of these pairs for the driving configurations $\xi^{1b},\xi^{2b}, \xi^{3b}$ and fleet sizes $\rho^\Phi = 0.1, 0.9$ in Figure \ref{fig:engineeringFD-flow} and Figure \ref{fig:engineeringFD-speed}. We recover the classical u-shape in the flow-occupancy plot. The blue points refer to the low volume traffic scenarios, the red points to the high volume traffic scenarios,  as introduced in Section~\ref{sec:sce}. This is also reflected by the fact that red points correspond to higher occupancy. Aggressiveness in driving decreases overall occupancy, increases speed, and increases flow, if the fleet is large. 

\begin{figure}

\begin{minipage}[t]{0.3\textwidth}
\centering
\includegraphics[scale=0.6]{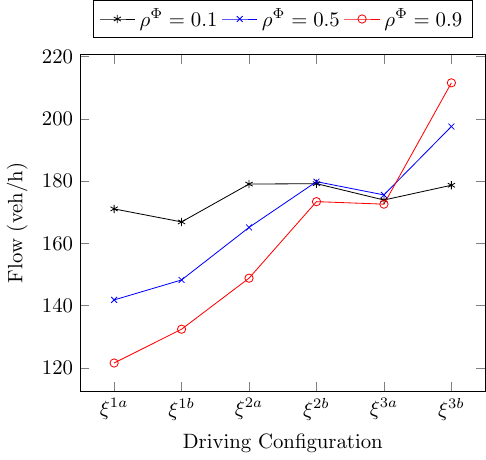}
\subcaption{Flow}
\end{minipage}\hspace{0.5cm}\begin{minipage}[t]{0.3\textwidth}
\centering
\includegraphics[scale=0.6]{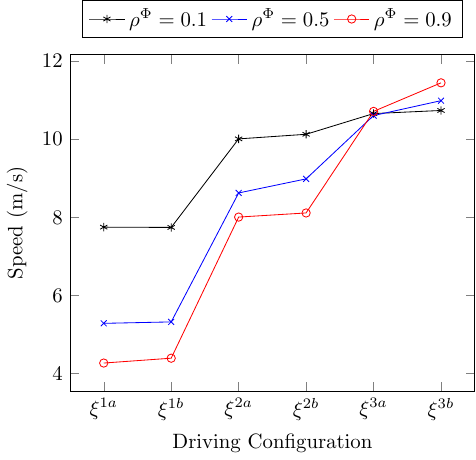}
\subcaption{Speed}
\end{minipage}\hspace{0.5cm}\begin{minipage}[t]{0.3\textwidth}
\centering
\includegraphics[scale=0.6]{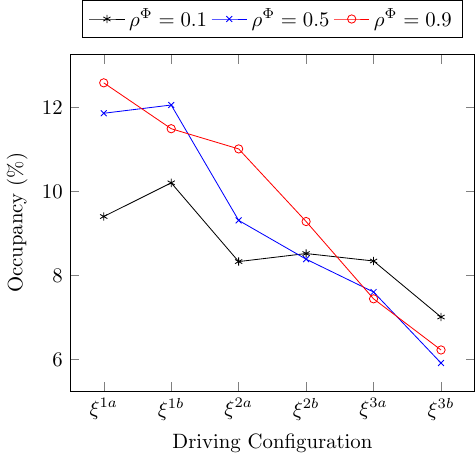}
\subcaption{Occupancy}
\end{minipage}
\caption{Impact of Fleet Size and Driving Configuration on Traffic System Performance.}
\label{fig:engineering2}
\end{figure}

To better understand the impact of individual driving configurations and fleet sizes, we study the scenario averages $\mathbb{E}\left( \sum_{k=1}^K \mu_k\bar{\chi}_k\right)=\frac{1}{K}\sum_{k=1}^K \bar{\chi}_k$. The results are displayed in Figure \ref{fig:engineering2}. We see that the performance of the traffic system improves with the aggressiveness of driving in the considered case studies; flows and average speeds increase, and the occupancy decreases.

\section{Supplementary Material: Number of Accidents}\label{sec:numberAccidents}

We present the expected number of accidents per scenario $\mathbb{E}(C^k)$, $k=1,\dots,K$, for both uniform and non-uniform accident occurrence. We also compute the expected total number of accidents $\mathbb{E}(\sum_{k=1^K} C^k)$. We note that, by definition, the involved expectations are equal for both Binomial and Poisson model and given by $\mathbb{E}(C^k)=N\mathbb{E}(\mu^k)p^k=N\mathbb{E}(\mu^k)\lambda^k$.\\

\textbf{Uniform Accident Occurrence.} For uniform accident occurrence, $p^k$ only depends on the fleet size $\rho^\Phi$. It neither depends on the scenario $k$, nor the driving configuration. This yields the values presented in Table \ref{tabAccUniform}.\\

\begin{table}[!htbp]
\centering 
\scalebox{1}{ 
\begin{threeparttable}
\caption[]{Expected Number of Accidents for Uniform Accident Occurrence.}
\label{tabAccUniform} 
\begin{tabular}{ccc} \toprule
\multicolumn{3}{c}{$\rho^\Phi$}\\ \midrule
 \quad\quad\quad 0.1\quad\quad\quad &\quad\quad\quad 0.5\quad\quad\quad & \quad\quad\quad 0.9 \quad\quad\quad \\ \toprule 
\quad\quad\quad 0.407 \quad\quad\quad & \quad\quad\quad 2.035 \quad\quad\quad & \quad\quad\quad 3.663 \quad\quad\quad\\ 
\bottomrule 
\end{tabular} 
\begin{tablenotes} 
\small 
\item The expected number of accidents $\mathbb{E}(C^k)=N\mathbb{E}(\mu^k)p^k$ is computed for the different fleet sizes with uniform accident occurrence.
\end{tablenotes} 
\end{threeparttable} 
}

\end{table}

\textbf{Non-Uniform Accident Occurrence.} For non-uniform accident occurrence, the expected number of accidents depends on the scenario, fleet size, and the respective driving configuration. The values for the different fleet sizes are presented in Table \ref{tabAcc01}, Table \ref{tabAcc05}, and Table \ref{tabAcc09}.\\

\begin{table} 
\centering 
\scalebox{0.8}{ 
\begin{threeparttable} 
\caption[]{Expected Number of Accidents for Non-Uniform Accident Occurrence and $\rho^\Phi=0.1$.} 
\label{tabAcc01}
\begin{tabular}{crrrrrrccrrrrrrr} \toprule 
Scenario &  $\xi^{1a}$ & $\xi^{1b}$ & $\xi^{2a}$ & $\xi^{2b}$ & $\xi^{3a}$ & $\xi^{3b}$ & \quad\quad &  Scenario &  $\xi^{1a}$ & $\xi^{1b}$ & $\xi^{2a}$ & $\xi^{2b}$ & $\xi^{3a}$ & $\xi^{3b}$ \\ \toprule 
       1 &    0.011 &    0.012 &    0.030 &    0.033 &    0.082 &    0.098 &   &    51 &    0.015 &    0.016 &    0.047 &    0.048 &    0.117 &    0.178\\ 
       2 &    0.023 &    0.021 &    0.064 &    0.063 &    0.161 &    0.239 &   &    52 &    0.031 &    0.031 &    0.100 &    0.092 &    0.216 &    0.270\\ 
       3 &    0.020 &    0.019 &    0.068 &    0.056 &    0.153 &    0.196 &   &    53 &    0.033 &    0.029 &    0.091 &    0.079 &    0.228 &    0.345\\ 
       4 &    0.017 &    0.014 &    0.073 &    0.067 &    0.179 &    0.214 &   &    54 &    0.029 &    0.026 &    0.109 &    0.111 &    0.294 &    0.352\\ 
       5 &    0.027 &    0.027 &    0.069 &    0.058 &    0.156 &    0.183 &   &    55 &    0.035 &    0.030 &    0.132 &    0.130 &    0.284 &    0.373\\ 
       6 &    0.018 &    0.021 &    0.062 &    0.062 &    0.162 &    0.158 &   &    56 &    0.047 &    0.046 &    0.120 &    0.132 &    0.349 &    0.442\\ 
       7 &    0.018 &    0.018 &    0.049 &    0.061 &    0.135 &    0.158 &   &    57 &    0.049 &    0.045 &    0.133 &    0.129 &    0.305 &    0.484\\ 
       8 &    0.042 &    0.038 &    0.046 &    0.046 &    0.137 &    0.135 &   &    58 &    0.063 &    0.057 &    0.151 &    0.157 &    0.405 &    0.565\\ 
       9 &    0.019 &    0.017 &    0.041 &    0.042 &    0.119 &    0.118 &   &    59 &    0.059 &    0.058 &    0.155 &    0.167 &    0.455 &    0.568\\ 
      10 &    0.025 &    0.023 &    0.059 &    0.040 &    0.107 &    0.147 &   &    60 &    0.091 &    0.083 &    0.187 &    0.156 &    0.573 &    0.572\\ 
      11 &    0.023 &    0.023 &    0.059 &    0.048 &    0.213 &    0.160 &   &    61 &    0.063 &    0.068 &    0.196 &    0.206 &    0.515 &    0.638\\ 
      12 &    0.020 &    0.014 &    0.056 &    0.043 &    0.118 &    0.129 &   &    62 &    0.061 &    0.064 &    0.210 &    0.183 &    0.567 &    0.577\\ 
      13 &    0.030 &    0.029 &    0.076 &    0.043 &    0.180 &    0.155 &   &    63 &    0.056 &    0.074 &    0.203 &    0.227 &    0.616 &    0.585\\ 
      14 &    0.026 &    0.026 &    0.073 &    0.041 &    0.152 &    0.134 &   &    64 &    0.054 &    0.062 &    0.220 &    0.200 &    0.693 &    0.655\\ 
      15 &    0.028 &    0.037 &    0.055 &    0.045 &    0.169 &    0.182 &   &    65 &    0.071 &    0.071 &    0.262 &    0.213 &    0.705 &    0.627\\ 
      16 &    0.049 &    0.043 &    0.069 &    0.050 &    0.210 &    0.233 &   &    66 &    0.083 &    0.088 &    0.250 &    0.191 &    0.621 &    0.538\\ 
      17 &    0.024 &    0.027 &    0.061 &    0.060 &    0.184 &    0.216 &   &    67 &    0.084 &    0.077 &    0.233 &    0.213 &    0.656 &    0.696\\ 
      18 &    0.031 &    0.030 &    0.082 &    0.051 &    0.184 &    0.222 &   &    68 &    0.081 &    0.113 &    0.232 &    0.248 &    0.625 &    0.688\\ 
      19 &    0.031 &    0.031 &    0.076 &    0.069 &    0.200 &    0.217 &   &    69 &    0.085 &    0.088 &    0.251 &    0.207 &    0.653 &    0.490\\ 
      20 &    0.028 &    0.034 &    0.098 &    0.108 &    0.176 &    0.228 &   &    70 &    0.069 &    0.094 &    0.274 &    0.278 &    0.650 &    0.798\\ 
      21 &    0.033 &    0.041 &    0.064 &    0.089 &    0.191 &    0.251 &   &    71 &    0.077 &    0.086 &    0.239 &    0.199 &    0.745 &    0.558\\ 
      22 &    0.024 &    0.023 &    0.085 &    0.075 &    0.223 &    0.243 &   &    72 &    0.098 &    0.101 &    0.277 &    0.278 &    0.644 &    0.595\\ 
      23 &    0.031 &    0.024 &    0.065 &    0.076 &    0.221 &    0.237 &   &    73 &    0.066 &    0.115 &    0.217 &    0.226 &    0.637 &    0.610\\ 
      24 &    0.037 &    0.033 &    0.084 &    0.076 &    0.185 &    0.207 &   &    74 &    0.071 &    0.105 &    0.241 &    0.251 &    0.789 &    0.676\\ 
      25 &    0.024 &    0.030 &    0.071 &    0.065 &    0.226 &    0.206 &   &    75 &    0.090 &    0.083 &    0.209 &    0.251 &    0.803 &    0.638\\ 
      26 &    0.031 &    0.030 &    0.068 &    0.067 &    0.162 &    0.144 &   &    76 &    0.102 &    0.079 &    0.249 &    0.223 &    0.672 &    0.575\\ 
      27 &    0.034 &    0.033 &    0.071 &    0.064 &    0.196 &    0.227 &   &    77 &    0.082 &    0.103 &    0.177 &    0.256 &    0.627 &    0.579\\ 
      28 &    0.027 &    0.033 &    0.083 &    0.071 &    0.216 &    0.188 &   &    78 &    0.091 &    0.074 &    0.256 &    0.198 &    0.774 &    0.497\\ 
      29 &    0.031 &    0.030 &    0.071 &    0.075 &    0.179 &    0.317 &   &    79 &    0.101 &    0.095 &    0.286 &    0.227 &    0.793 &    0.676\\ 
      30 &    0.031 &    0.024 &    0.064 &    0.078 &    0.108 &    0.223 &   &    80 &    0.090 &    0.072 &    0.226 &    0.223 &    0.782 &    0.605\\ 
      31 &    0.032 &    0.026 &    0.076 &    0.067 &    0.142 &    0.245 &   &    81 &    0.099 &    0.077 &    0.236 &    0.244 &    0.891 &    0.622\\ 
      32 &    0.029 &    0.028 &    0.088 &    0.080 &    0.181 &    0.260 &   &    82 &    0.115 &    0.085 &    0.231 &    0.257 &    0.772 &    0.454\\ 
      33 &    0.032 &    0.025 &    0.073 &    0.073 &    0.208 &    0.246 &   &    83 &    0.116 &    0.048 &    0.252 &    0.269 &    0.713 &    0.820\\ 
      34 &    0.026 &    0.031 &    0.097 &    0.072 &    0.204 &    0.217 &   &    84 &    0.094 &    0.077 &    0.207 &    0.225 &    0.807 &    0.738\\ 
      35 &    0.027 &    0.032 &    0.075 &    0.095 &    0.200 &    0.228 &   &    85 &    0.084 &    0.117 &    0.208 &    0.274 &    0.687 &    0.909\\ 
      36 &    0.039 &    0.036 &    0.071 &    0.067 &    0.221 &    0.223 &   &    86 &    0.076 &    0.098 &    0.203 &    0.167 &    0.682 &    0.601\\ 
      37 &    0.033 &    0.042 &    0.072 &    0.068 &    0.165 &    0.209 &   &    87 &    0.101 &    0.133 &    0.216 &    0.237 &    0.866 &    0.622\\ 
      38 &    0.026 &    0.026 &    0.069 &    0.054 &    0.197 &    0.232 &   &    88 &    0.104 &    0.116 &    0.271 &    0.268 &    0.623 &    0.513\\ 
      39 &    0.029 &    0.038 &    0.069 &    0.077 &    0.197 &    0.232 &   &    89 &    0.123 &    0.099 &    0.220 &    0.238 &    0.664 &    0.507\\ 
      40 &    0.028 &    0.031 &    0.072 &    0.071 &    0.210 &    0.231 &   &    90 &    0.124 &    0.075 &    0.192 &    0.213 &    0.582 &    0.603\\ 
      41 &    0.039 &    0.037 &    0.085 &    0.118 &    0.157 &    0.242 &   &    91 &    0.109 &    0.074 &    0.199 &    0.260 &    0.595 &    0.617\\ 
      42 &    0.023 &    0.027 &    0.063 &    0.108 &    0.218 &    0.251 &   &    92 &    0.139 &    0.085 &    0.233 &    0.231 &    0.711 &    0.607\\ 
      43 &    0.034 &    0.035 &    0.075 &    0.135 &    0.196 &    0.200 &   &    93 &    0.122 &    0.084 &    0.233 &    0.203 &    0.773 &    0.680\\ 
      44 &    0.030 &    0.046 &    0.099 &    0.137 &    0.220 &    0.261 &   &    94 &    0.103 &    0.098 &    0.267 &    0.271 &    0.745 &    0.767\\ 
      45 &    0.029 &    0.038 &    0.107 &    0.098 &    0.193 &    0.228 &   &    95 &    0.100 &    0.098 &    0.286 &    0.232 &    0.751 &    0.679\\ 
      46 &    0.031 &    0.041 &    0.139 &    0.104 &    0.203 &    0.225 &   &    96 &    0.088 &    0.100 &    0.283 &    0.226 &    0.642 &    0.736\\ 
      47 &    0.033 &    0.035 &    0.101 &    0.109 &    0.198 &    0.196 &   &    97 &    0.101 &    0.111 &    0.283 &    0.310 &    0.647 &    0.802\\ 
      48 &    0.044 &    0.061 &    0.105 &    0.102 &    0.172 &    0.218 &   &    98 &    0.081 &    0.095 &    0.314 &    0.325 &    0.638 &    0.691\\ 
      49 &    0.021 &    0.032 &    0.114 &    0.135 &    0.175 &    0.175 &   &    99 &    0.110 &    0.102 &    0.329 &    0.308 &    0.696 &    0.703\\ 
      50 &    0.016 &    0.028 &    0.087 &    0.077 &    0.159 &    0.120 &   &   100 &    0.050 &    0.081 &    0.168 &    0.257 &    0.500 &    0.494\\ 
 
\bottomrule 
\end{tabular} 
\begin{tablenotes} 
\small 
\item The expected number of accidents $\mathbb{E}(C^k)=N\mathbb{E}(\mu^k)p^k$ is computed for the different driving configurations with non-uniform accident occurrence and $\rho^\Phi=0.1$.
\end{tablenotes} 
\end{threeparttable} 
}

\end{table}

\begin{table} 
\centering 
\scalebox{0.8}{ 
\begin{threeparttable} 
\caption[]{Expected Number of Accidents for Non-Uniform Accident Occurrence and $\rho^\Phi=0.5$.} 
\label{tabAcc05}
\begin{tabular}{crrrrrrccrrrrrrr} \toprule 
Scenario &  $\xi^{1a}$ & $\xi^{1b}$ & $\xi^{2a}$ & $\xi^{2b}$ & $\xi^{3a}$ & $\xi^{3b}$ & \quad\quad &  Scenario &  $\xi^{1a}$ & $\xi^{1b}$ & $\xi^{2a}$ & $\xi^{2b}$ & $\xi^{3a}$ & $\xi^{3b}$ \\ \toprule 
       1 &    0.063 &    0.067 &    0.158 &    0.178 &    0.424 &    0.512 &   &    51 &    0.087 &    0.094 &    0.244 &    0.261 &    0.613 &    0.888\\ 
       2 &    0.128 &    0.126 &    0.287 &    0.312 &    0.926 &    1.236 &   &    52 &    0.165 &    0.172 &    0.512 &    0.509 &    1.277 &    1.791\\ 
       3 &    0.076 &    0.088 &    0.317 &    0.362 &    0.753 &    1.004 &   &    53 &    0.133 &    0.131 &    0.407 &    0.563 &    1.539 &    2.188\\ 
       4 &    0.070 &    0.063 &    0.340 &    0.343 &    0.957 &    1.228 &   &    54 &    0.149 &    0.141 &    0.526 &    0.650 &    1.617 &    1.980\\ 
       5 &    0.136 &    0.126 &    0.270 &    0.289 &    0.747 &    0.934 &   &    55 &    0.191 &    0.156 &    0.563 &    0.593 &    1.808 &    2.274\\ 
       6 &    0.078 &    0.102 &    0.264 &    0.284 &    0.971 &    0.930 &   &    56 &    0.218 &    0.231 &    0.522 &    0.639 &    1.959 &    2.337\\ 
       7 &    0.102 &    0.107 &    0.211 &    0.235 &    0.533 &    0.702 &   &    57 &    0.204 &    0.233 &    0.684 &    0.735 &    1.429 &    2.032\\ 
       8 &    0.119 &    0.122 &    0.225 &    0.194 &    0.561 &    0.705 &   &    58 &    0.223 &    0.173 &    0.561 &    0.652 &    1.922 &    2.272\\ 
       9 &    0.138 &    0.121 &    0.217 &    0.271 &    0.628 &    0.772 &   &    59 &    0.213 &    0.268 &    0.707 &    0.941 &    2.604 &    2.524\\ 
      10 &    0.130 &    0.123 &    0.299 &    0.251 &    0.606 &    0.999 &   &    60 &    0.240 &    0.214 &    0.784 &    0.858 &    2.953 &    2.798\\ 
      11 &    0.111 &    0.123 &    0.265 &    0.286 &    1.073 &    1.048 &   &    61 &    0.278 &    0.217 &    0.817 &    1.036 &    2.687 &    2.537\\ 
      12 &    0.122 &    0.133 &    0.211 &    0.294 &    1.050 &    0.819 &   &    62 &    0.272 &    0.349 &    0.898 &    0.958 &    2.813 &    3.100\\ 
      13 &    0.168 &    0.196 &    0.305 &    0.264 &    0.840 &    0.726 &   &    63 &    0.297 &    0.350 &    0.873 &    0.924 &    3.373 &    2.500\\ 
      14 &    0.158 &    0.124 &    0.205 &    0.214 &    0.980 &    0.789 &   &    64 &    0.404 &    0.346 &    0.840 &    1.063 &    2.989 &    2.898\\ 
      15 &    0.170 &    0.123 &    0.323 &    0.214 &    0.799 &    0.632 &   &    65 &    0.360 &    0.329 &    0.836 &    1.045 &    3.003 &    2.649\\ 
      16 &    0.148 &    0.134 &    0.352 &    0.243 &    1.009 &    0.621 &   &    66 &    0.319 &    0.384 &    0.963 &    1.025 &    2.705 &    3.167\\ 
      17 &    0.159 &    0.121 &    0.326 &    0.199 &    0.933 &    0.666 &   &    67 &    0.376 &    0.332 &    1.340 &    1.066 &    3.366 &    3.661\\ 
      18 &    0.206 &    0.164 &    0.365 &    0.315 &    1.006 &    0.945 &   &    68 &    0.411 &    0.292 &    1.028 &    1.117 &    3.318 &    3.275\\ 
      19 &    0.148 &    0.154 &    0.351 &    0.428 &    0.994 &    0.770 &   &    69 &    0.380 &    0.464 &    0.810 &    1.245 &    3.744 &    3.216\\ 
      20 &    0.180 &    0.110 &    0.360 &    0.362 &    0.912 &    0.929 &   &    70 &    0.513 &    0.483 &    1.062 &    1.217 &    2.970 &    3.024\\ 
      21 &    0.164 &    0.179 &    0.330 &    0.330 &    1.054 &    0.646 &   &    71 &    0.401 &    0.437 &    1.452 &    1.598 &    2.989 &    4.208\\ 
      22 &    0.189 &    0.139 &    0.490 &    0.409 &    1.001 &    0.918 &   &    72 &    0.391 &    0.401 &    1.032 &    1.269 &    2.794 &    3.274\\ 
      23 &    0.208 &    0.221 &    0.557 &    0.380 &    1.132 &    0.839 &   &    73 &    0.512 &    0.437 &    1.600 &    1.558 &    3.549 &    2.930\\ 
      24 &    0.158 &    0.178 &    0.429 &    0.278 &    0.755 &    0.672 &   &    74 &    0.336 &    0.520 &    1.319 &    1.372 &    2.913 &    3.336\\ 
      25 &    0.214 &    0.208 &    0.575 &    0.365 &    1.175 &    0.881 &   &    75 &    0.478 &    0.481 &    1.127 &    1.616 &    3.185 &    2.219\\ 
      26 &    0.233 &    0.224 &    0.507 &    0.420 &    0.798 &    0.777 &   &    76 &    0.410 &    0.561 &    1.271 &    1.367 &    2.591 &    2.516\\ 
      27 &    0.120 &    0.176 &    0.458 &    0.426 &    1.130 &    0.946 &   &    77 &    0.354 &    0.433 &    1.284 &    1.130 &    2.289 &    3.011\\ 
      28 &    0.263 &    0.199 &    0.549 &    0.317 &    1.305 &    1.108 &   &    78 &    0.551 &    0.416 &    1.220 &    0.890 &    2.830 &    2.836\\ 
      29 &    0.149 &    0.164 &    0.361 &    0.401 &    0.957 &    0.659 &   &    79 &    0.389 &    0.523 &    0.990 &    1.293 &    2.546 &    4.131\\ 
      30 &    0.158 &    0.190 &    0.353 &    0.337 &    0.960 &    0.763 &   &    80 &    0.357 &    0.552 &    1.213 &    1.126 &    2.879 &    3.642\\ 
      31 &    0.155 &    0.190 &    0.514 &    0.361 &    1.059 &    0.796 &   &    81 &    0.351 &    0.515 &    1.200 &    0.961 &    3.724 &    2.900\\ 
      32 &    0.154 &    0.215 &    0.683 &    0.385 &    0.611 &    0.711 &   &    82 &    0.390 &    0.428 &    1.186 &    1.323 &    3.836 &    3.237\\ 
      33 &    0.127 &    0.142 &    0.525 &    0.401 &    0.483 &    0.736 &   &    83 &    0.327 &    0.394 &    1.577 &    1.278 &    3.975 &    3.375\\ 
      34 &    0.100 &    0.172 &    0.571 &    0.384 &    0.873 &    0.810 &   &    84 &    0.371 &    0.427 &    1.633 &    1.169 &    3.878 &    3.206\\ 
      35 &    0.185 &    0.149 &    0.599 &    0.286 &    0.676 &    0.750 &   &    85 &    0.376 &    0.219 &    1.395 &    1.258 &    4.207 &    2.876\\ 
      36 &    0.177 &    0.206 &    0.521 &    0.416 &    0.980 &    0.681 &   &    86 &    0.433 &    0.352 &    1.616 &    0.914 &    3.997 &    3.670\\ 
      37 &    0.152 &    0.126 &    0.551 &    0.290 &    1.049 &    0.743 &   &    87 &    0.443 &    0.214 &    1.480 &    1.073 &    3.792 &    4.043\\ 
      38 &    0.160 &    0.174 &    0.446 &    0.349 &    0.922 &    0.694 &   &    88 &    0.403 &    0.253 &    1.307 &    0.911 &    2.677 &    4.091\\ 
      39 &    0.220 &    0.198 &    0.442 &    0.328 &    0.760 &    0.776 &   &    89 &    0.386 &    0.196 &    0.987 &    1.118 &    3.054 &    4.438\\ 
      40 &    0.139 &    0.150 &    0.444 &    0.333 &    0.937 &    0.730 &   &    90 &    0.374 &    0.408 &    1.080 &    1.298 &    3.068 &    3.832\\ 
      41 &    0.143 &    0.213 &    0.515 &    0.410 &    1.088 &    1.004 &   &    91 &    0.411 &    0.320 &    1.053 &    1.053 &    3.353 &    4.813\\ 
      42 &    0.135 &    0.147 &    0.624 &    0.482 &    0.961 &    0.744 &   &    92 &    0.374 &    0.481 &    1.105 &    1.343 &    3.303 &    5.189\\ 
      43 &    0.139 &    0.150 &    0.494 &    0.723 &    0.958 &    0.926 &   &    93 &    0.380 &    0.496 &    1.243 &    1.317 &    3.400 &    4.766\\ 
      44 &    0.169 &    0.162 &    0.441 &    0.539 &    0.922 &    0.887 &   &    94 &    0.408 &    0.450 &    1.172 &    1.382 &    4.417 &    5.476\\ 
      45 &    0.227 &    0.180 &    0.584 &    0.489 &    1.168 &    0.837 &   &    95 &    0.265 &    0.432 &    1.132 &    1.571 &    4.071 &    4.528\\ 
      46 &    0.122 &    0.127 &    0.481 &    0.320 &    1.455 &    0.679 &   &    96 &    0.274 &    0.440 &    1.029 &    1.450 &    3.714 &    4.625\\ 
      47 &    0.186 &    0.159 &    0.519 &    0.490 &    0.990 &    0.740 &   &    97 &    0.416 &    0.391 &    0.776 &    1.287 &    3.569 &    3.219\\ 
      48 &    0.253 &    0.164 &    0.460 &    0.447 &    0.852 &    0.716 &   &    98 &    0.276 &    0.584 &    1.393 &    1.487 &    3.659 &    3.135\\ 
      49 &    0.155 &    0.138 &    0.455 &    0.476 &    0.804 &    0.954 &   &    99 &    0.555 &    0.467 &    1.267 &    1.379 &    3.765 &    2.717\\ 
      50 &    0.077 &    0.066 &    0.301 &    0.292 &    0.563 &    0.238 &   &   100 &    0.145 &    0.157 &    0.598 &    0.986 &    1.552 &    1.396\\

\bottomrule 
\end{tabular} 
\begin{tablenotes} 
\small 
\item The expected number of accidents $\mathbb{E}(C^k)=N\mathbb{E}(\mu^k)p^k$ is computed for the different driving configurations with non-uniform accident occurrence and $\rho^\Phi=0.5$.
\end{tablenotes} 
\end{threeparttable} 
}

\end{table}

\begin{table} 
\centering 
\scalebox{0.8}{ 
\begin{threeparttable} 
\caption[]{Expected Number of Accidents for Non-Uniform Accident Occurrence and $\rho^\Phi=0.9$.} 
\label{tabAcc09}
\begin{tabular}{crrrrrrccrrrrrrr} \toprule 
Scenario &  $\xi^{1a}$ & $\xi^{1b}$ & $\xi^{2a}$ & $\xi^{2b}$ & $\xi^{3a}$ & $\xi^{3b}$ & \quad\quad &  Scenario &  $\xi^{1a}$ & $\xi^{1b}$ & $\xi^{2a}$ & $\xi^{2b}$ & $\xi^{3a}$ & $\xi^{3b}$ \\ \toprule 
       1 &    0.210 &    0.248 &    0.404 &    0.437 &    0.914 &    0.990 &   &    51 &    0.260 &    0.298 &    0.535 &    0.667 &    1.451 &    1.711\\ 
       2 &    0.287 &    0.314 &    0.477 &    0.530 &    1.549 &    1.770 &   &    52 &    0.406 &    0.439 &    0.762 &    0.868 &    2.164 &    3.144\\ 
       3 &    0.163 &    0.166 &    0.544 &    0.583 &    1.633 &    1.527 &   &    53 &    0.376 &    0.382 &    0.754 &    0.957 &    2.538 &    3.412\\ 
       4 &    0.143 &    0.190 &    0.591 &    0.668 &    1.814 &    1.801 &   &    54 &    0.301 &    0.322 &    0.998 &    1.153 &    2.934 &    3.744\\ 
       5 &    0.226 &    0.246 &    0.471 &    0.613 &    1.449 &    1.896 &   &    55 &    0.259 &    0.450 &    0.784 &    1.321 &    2.952 &    4.352\\ 
       6 &    0.309 &    0.285 &    0.406 &    0.539 &    1.652 &    1.492 &   &    56 &    0.423 &    0.486 &    1.059 &    1.127 &    3.487 &    3.803\\ 
       7 &    0.242 &    0.257 &    0.493 &    0.508 &    1.580 &    1.517 &   &    57 &    0.535 &    0.482 &    0.872 &    1.300 &    3.307 &    4.277\\ 
       8 &    0.251 &    0.224 &    0.393 &    0.393 &    1.250 &    1.218 &   &    58 &    0.424 &    0.446 &    1.226 &    1.110 &    3.606 &    4.653\\ 
       9 &    0.263 &    0.270 &    0.410 &    0.402 &    1.139 &    1.274 &   &    59 &    0.371 &    0.434 &    1.061 &    1.428 &    5.064 &    5.213\\ 
      10 &    0.277 &    0.283 &    0.485 &    0.390 &    1.280 &    1.149 &   &    60 &    0.596 &    0.531 &    1.130 &    1.416 &    4.542 &    4.558\\ 
      11 &    0.251 &    0.250 &    0.690 &    0.485 &    2.027 &    1.528 &   &    61 &    0.371 &    0.478 &    1.172 &    1.469 &    5.310 &    6.026\\ 
      12 &    0.232 &    0.396 &    0.643 &    0.438 &    1.789 &    1.191 &   &    62 &    0.478 &    0.453 &    1.431 &    1.583 &    5.881 &    6.296\\ 
      13 &    0.244 &    0.287 &    0.553 &    0.474 &    2.114 &    1.350 &   &    63 &    0.493 &    0.500 &    1.762 &    1.832 &    6.149 &    5.773\\ 
      14 &    0.296 &    0.327 &    0.668 &    0.440 &    2.016 &    1.140 &   &    64 &    0.471 &    0.522 &    1.392 &    2.168 &    6.499 &    6.421\\ 
      15 &    0.402 &    0.387 &    0.497 &    0.365 &    1.706 &    1.208 &   &    65 &    0.518 &    0.481 &    1.697 &    2.105 &    6.589 &    5.943\\ 
      16 &    0.327 &    0.313 &    0.867 &    0.469 &    1.834 &    1.105 &   &    66 &    0.590 &    0.608 &    1.666 &    1.725 &    5.099 &    6.180\\ 
      17 &    0.380 &    0.271 &    0.613 &    0.467 &    2.007 &    1.234 &   &    67 &    0.720 &    0.572 &    1.969 &    1.838 &    5.481 &    5.559\\ 
      18 &    0.215 &    0.270 &    0.641 &    0.434 &    1.632 &    1.260 &   &    68 &    0.709 &    0.439 &    1.681 &    2.332 &    5.331 &    7.396\\ 
      19 &    0.227 &    0.264 &    0.581 &    0.386 &    2.032 &    1.187 &   &    69 &    0.620 &    0.506 &    1.937 &    1.692 &    5.138 &    6.560\\ 
      20 &    0.416 &    0.365 &    0.657 &    0.496 &    2.012 &    1.342 &   &    70 &    0.559 &    0.600 &    1.447 &    2.172 &    5.150 &    6.256\\ 
      21 &    0.220 &    0.320 &    0.572 &    0.402 &    2.080 &    1.366 &   &    71 &    0.648 &    0.593 &    1.776 &    2.041 &    6.764 &    7.774\\ 
      22 &    0.260 &    0.286 &    0.695 &    0.531 &    2.215 &    1.675 &   &    72 &    0.744 &    0.612 &    2.249 &    1.709 &    6.999 &    5.912\\ 
      23 &    0.271 &    0.342 &    0.895 &    0.738 &    2.302 &    1.458 &   &    73 &    0.738 &    0.831 &    1.805 &    1.920 &    6.551 &    6.062\\ 
      24 &    0.280 &    0.400 &    0.620 &    0.715 &    1.869 &    1.230 &   &    74 &    0.669 &    0.615 &    2.074 &    2.042 &    6.351 &    5.598\\ 
      25 &    0.239 &    0.350 &    0.662 &    0.699 &    1.993 &    1.370 &   &    75 &    0.574 &    0.621 &    1.762 &    2.243 &    4.736 &    7.480\\ 
      26 &    0.260 &    0.496 &    0.725 &    0.436 &    1.753 &    1.155 &   &    76 &    0.588 &    0.614 &    1.858 &    2.150 &    6.260 &    7.337\\ 
      27 &    0.283 &    0.332 &    0.565 &    0.540 &    2.034 &    1.317 &   &    77 &    0.529 &    0.476 &    2.130 &    1.893 &    5.756 &    5.461\\ 
      28 &    0.260 &    0.324 &    0.898 &    0.529 &    1.974 &    1.049 &   &    78 &    0.507 &    0.560 &    1.881 &    2.278 &    4.603 &    5.212\\ 
      29 &    0.495 &    0.294 &    0.662 &    0.570 &    1.873 &    1.331 &   &    79 &    0.479 &    0.633 &    1.965 &    2.036 &    4.229 &    5.506\\ 
      30 &    0.350 &    0.289 &    0.963 &    0.433 &    2.010 &    1.215 &   &    80 &    0.491 &    0.754 &    1.414 &    2.478 &    4.102 &    6.615\\ 
      31 &    0.398 &    0.268 &    0.574 &    0.632 &    1.641 &    1.359 &   &    81 &    0.366 &    0.692 &    2.418 &    2.221 &    4.906 &    4.359\\ 
      32 &    0.321 &    0.314 &    0.438 &    0.847 &    2.325 &    1.564 &   &    82 &    0.585 &    0.739 &    2.081 &    2.322 &    5.959 &    6.304\\ 
      33 &    0.338 &    0.272 &    0.428 &    1.309 &    1.897 &    1.637 &   &    83 &    0.592 &    0.627 &    2.665 &    1.675 &    4.790 &    6.161\\ 
      34 &    0.369 &    0.270 &    0.286 &    0.623 &    2.006 &    1.366 &   &    84 &    0.680 &    0.673 &    1.780 &    2.044 &    5.923 &    6.777\\ 
      35 &    0.466 &    0.456 &    0.467 &    0.887 &    1.964 &    1.387 &   &    85 &    0.619 &    0.513 &    1.458 &    1.512 &    7.145 &    6.888\\ 
      36 &    0.371 &    0.293 &    0.754 &    0.640 &    2.163 &    1.318 &   &    86 &    0.414 &    0.614 &    1.707 &    1.731 &    6.458 &    7.630\\ 
      37 &    0.411 &    0.233 &    0.684 &    0.681 &    1.902 &    1.182 &   &    87 &    0.618 &    0.575 &    1.713 &    1.955 &    7.100 &    5.358\\ 
      38 &    0.398 &    0.281 &    0.678 &    0.743 &    1.838 &    1.249 &   &    88 &    0.462 &    0.490 &    2.347 &    1.834 &    6.681 &    7.801\\ 
      39 &    0.396 &    0.269 &    0.558 &    0.528 &    1.883 &    1.513 &   &    89 &    0.416 &    0.680 &    2.425 &    2.981 &    6.499 &    4.090\\ 
      40 &    0.524 &    0.375 &    0.741 &    0.751 &    1.979 &    1.452 &   &    90 &    0.499 &    0.629 &    1.539 &    3.613 &    5.169 &    5.765\\ 
      41 &    0.362 &    0.284 &    0.772 &    0.667 &    2.446 &    1.071 &   &    91 &    0.648 &    0.644 &    1.859 &    3.113 &    4.926 &    4.817\\ 
      42 &    0.441 &    0.310 &    0.857 &    0.757 &    1.996 &    1.653 &   &    92 &    0.458 &    0.657 &    2.884 &    3.438 &    6.093 &    4.222\\ 
      43 &    0.457 &    0.303 &    0.919 &    0.975 &    1.944 &    1.471 &   &    93 &    0.530 &    0.700 &    3.324 &    2.417 &    5.460 &    5.895\\ 
      44 &    0.463 &    0.461 &    0.829 &    0.640 &    1.966 &    1.383 &   &    94 &    0.829 &    0.780 &    3.284 &    2.561 &    5.435 &    5.650\\ 
      45 &    0.666 &    0.340 &    0.591 &    0.702 &    2.016 &    1.261 &   &    95 &    0.446 &    0.766 &    2.211 &    2.420 &    5.502 &    6.146\\ 
      46 &    0.683 &    0.596 &    0.826 &    0.617 &    2.132 &    1.203 &   &    96 &    0.822 &    0.641 &    2.095 &    1.994 &    4.905 &    6.230\\ 
      47 &    0.690 &    0.650 &    0.511 &    0.777 &    1.819 &    1.264 &   &    97 &    0.708 &    0.504 &    2.410 &    1.986 &    5.911 &    6.273\\ 
      48 &    0.543 &    0.338 &    0.676 &    0.584 &    1.874 &    1.120 &   &    98 &    0.425 &    0.428 &    1.741 &    2.001 &    5.398 &    6.954\\ 
      49 &    0.449 &    0.319 &    0.523 &    0.722 &    1.743 &    1.264 &   &    99 &    0.533 &    0.512 &    2.714 &    2.325 &    6.597 &    7.637\\ 
      50 &    0.245 &    0.164 &    0.488 &    0.466 &    1.586 &    0.810 &   &   100 &    0.213 &    0.234 &    1.043 &    1.066 &    4.294 &    4.861\\ 

\bottomrule 
\end{tabular} 
\begin{tablenotes} 
\small 
\item The expected number of accidents $\mathbb{E}(C^k)=N\mathbb{E}(\mu^k)p^k$ is computed for the different driving configurations with non-uniform accident occurrence and $\rho^\Phi=0.9$.
\end{tablenotes} 
\end{threeparttable} 
}

\end{table}

\textbf{Total Number of Accidents.} We contrast the expected total number of accidents for uniform and non-uniform accident occurrence in Table \ref{accTotal}.

\begin{table}[!htbp]
\centering 
\scalebox{0.8}{ 
\begin{threeparttable} 
\caption[]{Expected Total Number of Accidents.} 
\label{accTotal}
\begin{tabular}{rcc} \toprule 
&  Uniform  & Non-Uniform    \\     \toprule 
 \\ 
$\mathbf{\rho^\Phi=0.1:}$   &       & \\ 
$\xi^{1a}$ & 40.7 & 5.5 \\ 
$\xi^{1b}$ & 40.7 & 5.5 \\ 
$\xi^{2a}$ & 40.7 & 14.5 \\ 
$\xi^{2b}$ & 40.7 & 14.4 \\ 
$\xi^{3a}$ & 40.7 & 39.7\\ 
$\xi^{3b}$ & 40.7 & 39.8\\ 
$\mathbf{\rho^\Phi=0.5:}$   &       & \\ 
$\xi^{1a}$ & 203.5 & 24.6 \\ 
$\xi^{1b}$ & 203.5 & 25.3 \\ 
$\xi^{2a}$ & 203.5 & 72.1 \\ 
$\xi^{2b}$ & 203.5 & 72.3 \\ 
$\xi^{3a}$ & 203.5 & 193.3 \\ 
$\xi^{3b}$ & 203.5 & 199.1 \\ 
$\mathbf{\rho^\Phi=0.9:}$   &       &  \\ 
$\xi^{1a}$ & 366.3 & 43.6  \\ 
$\xi^{1b}$ & 366.3 & 43.7 \\ 
$\xi^{2a}$ & 366.3 & 118.9 \\ 
$\xi^{2b}$ & 366.3 & 125.9 \\ 
$\xi^{3a}$ & 366.3 & 352.8 \\ 
$\xi^{3b}$ & 366.3 & 351.0 \\  
\bottomrule 
\end{tabular} 
\begin{tablenotes} 
\small 
\item The expected total number of accidents is given by $\mathbb{E}(\sum_{k=1}^K C^k)$ and computed for uniform and non-uniform accident occurrence. 
\end{tablenotes} 
\end{threeparttable} 
}

\end{table}

\section{Supplementary Material: Sampling Procedure}\label{sec:sampling}

We provide a detailed pseudo-code for the procedure to obtain samples from $L$ in Algorithm \ref{alg:lossSampling}. In our case studies, we use $M'=M=10,000$ samples of $\psi$ in each scenario $k$ to approximate its distribution $\lcal^{k}$. We note that, instead of this bootstrapping approach, one could also pre-sample sufficiently many values.

\begin{algorithm}[!htbp]

\caption{Sampling of Losses $L$.}
\label{alg:lossSampling}

\footnotesize

\begin{algorithmic}
\STATE{\textbf{Phase 1:}} Prior Evaluation of Traffic Model
\STATE{}
\FOR{$k=1,\dots,K$}
\STATE{Run SUMO in scenario $k$.}
\STATE{Obtain data to calculate $p^k_r$ and $\lambda^k_r$ for $r=1,\dots,R$.}
\STATE{Terminate SUMO.}
\STATE{Set $p^k=\sum_{r=1}^R p^k_r$ and $\lambda^k=\sum_{r=1}^R \lambda^k_r$.}
\ENDFOR
\STATE{}
\STATE{\textbf{Phase 2:}} Pre-Sampling of $\psi$ conditional on scenario $k=1,\dots,K$.
\STATE{}
\FOR{$k=1,\dots,K$}
\FOR{$j=1,\dots,M'$}
\STATE{Sample $\hat{t}_j=\mathrm{Unif}(0,T)$.}
\STATE{Sample $\hat{r}_j\sim \mathcal{R}$ where $P(\mathcal{R}=r)=p_r^k/p^k$ (or $P(\mathcal{R}=r)=\lambda_r^k/\lambda^k$).}
\ENDFOR
\STATE{Sort $\hat{t}_1,\dots,\hat{t}_{M'}$ by size (again denoted by $\hat{t}_1,\dots,\hat{t}_{M'}$).}
\STATE{Start SUMO in scenario $k$.}
\FOR{$j=1,\dots,M'$}
\STATE{Continue SUMO until time $\hat{t}_{j}$.}
\STATE{Sample $\hat{i}_{j}\sim\mathrm{Unif}\left(\mathcal{M}^\Phi_{\hat{r}_{j}}\left(\hat{t}_{j}\right)\right)$.}
\STATE{Set $\hat{\psi}^k_{j}=v^{\hat{i}_{j}}(\hat{t}_{j})$.}
\ENDFOR
\STATE{Terminate SUMO.}
\STATE{Store $\hat{\psi}^k_1,\dots,\hat{\psi}^k_{M'}$.}
\ENDFOR
\STATE{}
\STATE{\textbf{Phase 3:}} Sampling of total losses $L$.
\STATE{}
\FOR{$j=1,\dots,M$}
\STATE{Sample $\hat{y}^j$ with $P(y=g)=P(y=b)=1/2$.}
\FOR{$n=1,\dots,N$}
\STATE{Sample $\hat{\nu}^{j,n}\sim \nu_{\hat{y}^j}$.}
\ENDFOR
\FOR{$k=1,\dots,K$}
\STATE{Set $(\hat{\mu}^k)^j=1/N\sum_{n=1}^N \mathbbm{1}\{\hat{\nu}^{j,n}=k\}$.}
\ENDFOR
\FOR{$k=1,\dots,K$}
\STATE{Sample $\left(\hat{C}^k\right)^j\sim \mathrm{Bin}(p^k,N(\hat{\mu}^k)^j)$ (or $\left(\hat{C}^k\right)^j\sim \mathrm{Poiss}(\lambda^kN(\hat{\mu}^k)^j)$).}
\FOR{$c=1,\dots,\left(\hat{C}^k\right)^j$}
\STATE{Sample $\hat{l}\sim\mathrm{Unif}(\{1,\dots,M'\})$.}
\STATE{Sample $\hat{X}^k_{c}\sim  F^{\hat{\psi}^k_{\hat{l}}}$.}
\ENDFOR
\ENDFOR
\STATE{Set $\hat{L}^j=\sum_{k=1}^K \sum_{c=1}^{\left(\hat{C}^k\right)^j} \hat{X}^{k}_c$.}
\ENDFOR
\STATE{}
\STATE{\textbf{Output:} $\hat{L}^1,\dots,\hat{L}^{M}$.}
\end{algorithmic}

\end{algorithm}

\section{Supplementary Material: Tables}\label{sec:tables}

We provide detailed tables that contain selected statistical functionals of the total loss in our different case studies. We evaluate
\begin{enumerate}
\item \emph{Expectation.} $\mathbb{E}(L)$, 
\item \emph{Variance.} $\mathrm{Var}(L)=\mathbb{E}(L^2)-\mathbb{E}(L)^2$,
\item \emph{Skewness.} $\varsigma_{L}=\frac{\mathbb{E}\left[(L-\mathbb{E}(L))^3\right]}{(\mathrm{Var}(L))^{3/2}}$,
\item \emph{Value-at-Risk.} $\mathrm{VaR}_{p}(L)=\inf \{x\in \mathbb{R}\colon P(L\leq x)\geq p\}$,~$p=0.9,~0.95,~0.99$,
\item \emph{Expected Shortfall.} $\mathrm{ES}_{p}(L)=\frac{1}{1-p}\int_p^1 \mathrm{VaR}_q(L)\,\mathrm{d}q$,~$p=0.9,~0.95,~0.99$.
\end{enumerate}
These statistical functionals are presented for both unnormalized and normalized values of $L$. In total, we provide 18 tables, as shown in the list of tables below: The first 9 tables contain the statistical functionals for unnormalized total losses while the second 9 tables contain the results for normalized total losses.

\setcounter{table}{0}
\listoftables
~\\

\textbf{Remarks on the Normalization.} Recall that we normalize total losses in order to compare losses among fleet sizes $\rho^\Phi=0.1,~0.5,~0.9$. More precisely, we normalize $L$ by \emph{100 expected insured vehicles} as follows:
\begin{itemize}
\item In the underlying SUMO scneario, the route files specify a number of vehicles for each \emph{flow} belonging to the fleet $\Phi$. 
\item For each traffic scenario $k$, we denote the sum of these values over all flows by $n^{k}$. We interpret this as the total number of insured vehicles in traffic scenario $k$. In our case studies, $n^{k}$ takes two different values corresponding to the good and bad scenarios:
\begin{equation*}
n^{k}=\begin{cases}
n_{g}, & k=1,\dots, 50,\\
n_{b}, & k=51,\dots, 100.
\end{cases}
\end{equation*}
\item We denote the total number of insured vehicles by $n^{\Phi}$. It is given by $n^{\Phi} = \sum_{k=1}^K \mu^kn^{k}$. Note that this number is random as $\mu$ is random.
\item We evaluate the normalized total loss per $100$ expected insured vehicles. According to our specific choice of $\mu$, it is given by
\begin{align*}
\frac{L}{\frac{\mathbb{E}(n^{\Phi})}{100}}&= 100\cdot \frac{L}{\sum_{k=1}^K n^{k}\mathbb{E}(\mu^k)}=\frac{200}{n_{g}+n_{b}}\cdot L.
\end{align*}
\end{itemize}

\begin{table} 
\centering 
\scalebox{0.7}{ 
\begin{threeparttable} 
\caption{Expectation of the Total Loss.} 
\begin{tabular}{rrrrrrrrrrrrrr} \toprule 
& \multicolumn{6}{c}{Binomial Model}  & \quad\quad & \multicolumn{6}{c}{Poisson Model}   \\ \midrule 
& \multicolumn{3}{c}{Gamma}  & \multicolumn{3}{c}{Log-Normal} & & \multicolumn{3}{c}{Gamma}  & \multicolumn{3}{c}{Log-Normal}  \\ \midrule 
& $c_v=0.5$ & $c_v=1.0$ & $c_v=2.0$  & $c_v=0.5$ & $c_v=1.0$ & $c_v=2.0$ & & $c_v=0.5$ & $c_v=1.0$ & $c_v=2.0$  &  $c_v=0.5$ & $c_v=1.0$ & $c_v=2.0$  \\     \toprule 
 \\ 
&\multicolumn{13}{c}{\textbf{Uniform Accident Occurrence}}  \\ 
$\mathbf{\rho^\Phi=0.1:}$   &       & &   &     &  & & &       &   &  & & & \\ 
$\xi^{1a}$ &    546.7 &    544.7 &    544.2 &    547.0 &    544.2 &    546.5 & &    545.5 &    546.7 &    546.2 &    543.4 &    547.2 &    542.5 \\ 
$\xi^{1b}$ &    527.0 &    525.9 &    523.0 &    529.3 &    524.7 &    522.1 & &    527.0 &    527.4 &    523.0 &    523.2 &    529.5 &    525.7 \\
\bottomrule 
\end{tabular} 
\begin{tablenotes} 
\small 
\item The expectation of the total loss is approximated using $10,000$ independent samples of $L$. 
\end{tablenotes} 
\end{threeparttable} 
}
\label{tab1}

\end{table}

\begin{table} 
\centering 
\scalebox{0.57}{ 
\begin{threeparttable} 
\caption{Variance of the Total Loss.} 
\begin{tabular}{rrrrrrrrrrrrrr} \toprule 
& \multicolumn{6}{c}{Binomial Model}  & \quad\quad & \multicolumn{6}{c}{Poisson Model}   \\ \midrule 
& \multicolumn{3}{c}{Gamma}  & \multicolumn{3}{c}{Log-Normal} & & \multicolumn{3}{c}{Gamma}  & \multicolumn{3}{c}{Log-Normal}  \\ \midrule 
& $c_v=0.5$ & $c_v=1.0$ & $c_v=2.0$  & $c_v=0.5$ & $c_v=1.0$ & $c_v=2.0$ & & $c_v=0.5$ & $c_v=1.0$ & $c_v=2.0$  &  $c_v=0.5$ & $c_v=1.0$ & $c_v=2.0$  \\     \toprule 
 \\ 
&\multicolumn{13}{c}{\textbf{Uniform Accident Occurrence}}  \\ 
$\mathbf{\rho^\Phi=0.1:}$   &       & &   &     &  & & &       &   &  & & & \\ 
$\xi^{1a}$ &  18893.7 &  27665.8 &  63330.2 &  18797.9 &  28026.1 &  65488.9 & &  18277.8 &  28087.5 &  63245.1 &  19071.7 &  28092.0 &  62943.3 \\ 
$\xi^{1b}$ &  18087.2 &  26745.3 &  60657.5 &  18440.3 &  26108.7 &  61767.7 & &  17887.3 &  26811.1 &  60802.6 &  17589.5 &  27311.8 &  61022.4 \\
\bottomrule 
\end{tabular} 
\begin{tablenotes} 
\small 
\item The variance of the total loss is approximated using $10,000$ independent samples of $L$. 
\end{tablenotes} 
\end{threeparttable} 
}
\label{tab2}

\end{table}

\begin{table} 
\centering 
\scalebox{0.7}{ 
\begin{threeparttable} 
\caption{Skewness of the Total Loss.} 
\begin{tabular}{rrrrrrrrrrrrrr} \toprule 
& \multicolumn{6}{c}{Binomial Model}  & \quad\quad & \multicolumn{6}{c}{Poisson Model}   \\ \midrule 
& \multicolumn{3}{c}{Gamma}  & \multicolumn{3}{c}{Log-Normal} & & \multicolumn{3}{c}{Gamma}  & \multicolumn{3}{c}{Log-Normal}  \\ \midrule 
& $c_v=0.5$ & $c_v=1.0$ & $c_v=2.0$  & $c_v=0.5$ & $c_v=1.0$ & $c_v=2.0$ & & $c_v=0.5$ & $c_v=1.0$ & $c_v=2.0$  &  $c_v=0.5$ & $c_v=1.0$ & $c_v=2.0$  \\     \toprule 
 \\ 
&\multicolumn{13}{c}{\textbf{Uniform Accident Occurrence}}  \\ 
$\mathbf{\rho^\Phi=0.1:}$   &       & &   &     &  & & &       &   &  & & & \\ 
$\xi^{1a}$ &    0.306 &    0.533 &    0.869 &    0.320 &    0.679 &    1.904 & &    0.327 &    0.454 &    0.858 &    0.386 &    0.536 &    1.728 \\ 
$\xi^{1b}$ &    0.292 &    0.468 &    0.868 &    0.357 &    0.666 &    1.764 & &    0.330 &    0.468 &    0.818 &    0.366 &    0.563 &    1.756 \\
\bottomrule 
\end{tabular} 
\begin{tablenotes} 
\small 
\item The skewness of the total loss is approximated using $10,000$ independent samples of $L$. 
\end{tablenotes} 
\end{threeparttable} 
}
\label{tab3}

\end{table}

\begin{table} 
\centering 
\scalebox{0.7}{ 
\begin{threeparttable} 
\caption{Value-at-Risk at Level $p=0.9$ of the Total Loss.} 
\begin{tabular}{rrrrrrrrrrrrrr} \toprule 
& \multicolumn{6}{c}{Binomial Model}  & \quad\quad & \multicolumn{6}{c}{Poisson Model}   \\ \midrule 
& \multicolumn{3}{c}{Gamma}  & \multicolumn{3}{c}{Log-Normal} & & \multicolumn{3}{c}{Gamma}  & \multicolumn{3}{c}{Log-Normal}  \\ \midrule 
& $c_v=0.5$ & $c_v=1.0$ & $c_v=2.0$  & $c_v=0.5$ & $c_v=1.0$ & $c_v=2.0$ & & $c_v=0.5$ & $c_v=1.0$ & $c_v=2.0$  &  $c_v=0.5$ & $c_v=1.0$ & $c_v=2.0$  \\     \toprule 
 \\ 
&\multicolumn{13}{c}{\textbf{Uniform Accident Occurrence}}  \\ 
$\mathbf{\rho^\Phi=0.1:}$   &       & &   &     &  & & &       &   &  & & & \\ 
$\xi^{1a}$ &    726.8 &    766.2 &    882.9 &    727.7 &    763.1 &    861.4 & &    725.3 &    772.8 &    887.0 &    723.0 &    771.9 &    851.7 \\ 
$\xi^{1b}$ &    707.7 &    743.7 &    848.1 &    709.3 &    737.7 &    829.6 & &    706.0 &    743.0 &    857.3 &    699.0 &    748.5 &    827.4 \\
\bottomrule 
\end{tabular} 
\begin{tablenotes} 
\small 
\item The Value-at-Risk at level $p=0.9$ of the total loss is approximated using $10,000$ independent samples of $L$. 
\end{tablenotes} 
\end{threeparttable} 
}
\label{tab4}

\end{table}

\begin{table} 
\centering 
\scalebox{0.7}{ 
\begin{threeparttable} 
\caption{Expected Shortfall at Level $p=0.9$ of the Total Loss.} 
\begin{tabular}{rrrrrrrrrrrrrr} \toprule 
& \multicolumn{6}{c}{Binomial Model}  & \quad\quad & \multicolumn{6}{c}{Poisson Model}   \\ \midrule 
& \multicolumn{3}{c}{Gamma}  & \multicolumn{3}{c}{Log-Normal} & & \multicolumn{3}{c}{Gamma}  & \multicolumn{3}{c}{Log-Normal}  \\ \midrule 
& $c_v=0.5$ & $c_v=1.0$ & $c_v=2.0$  & $c_v=0.5$ & $c_v=1.0$ & $c_v=2.0$ & & $c_v=0.5$ & $c_v=1.0$ & $c_v=2.0$  &  $c_v=0.5$ & $c_v=1.0$ & $c_v=2.0$  \\     \toprule 
 \\ 
&\multicolumn{13}{c}{\textbf{Uniform Accident Occurrence}}  \\ 
$\mathbf{\rho^\Phi=0.1:}$   &       & &   &     &  & & &       &   &  & & & \\ 
$\xi^{1a}$ &    801.9 &    866.5 &   1055.4 &    803.7 &    871.6 &   1096.5 & &    797.8 &    868.0 &   1058.7 &    805.2 &    875.7 &   1079.0 \\ 
$\xi^{1b}$ &    778.2 &    840.5 &   1025.3 &    786.6 &    844.8 &   1058.5 & &    777.6 &    842.5 &   1022.9 &    773.4 &    853.6 &   1054.4 \\
\bottomrule 
\end{tabular} 
\begin{tablenotes} 
\small 
\item The Expected Shortfall at level $p=0.9$ of the total loss is approximated using $10,000$ independent samples of $L$. 
\end{tablenotes} 
\end{threeparttable} 
}
\label{tab5}

\end{table}

\begin{table} 
\centering 
\scalebox{0.7}{ 
\begin{threeparttable} 
\caption{Value-at-Risk at Level $p=0.95$ of the Total Loss.} 
\begin{tabular}{rrrrrrrrrrrrrr} \toprule 
& \multicolumn{6}{c}{Binomial Model}  & \quad\quad & \multicolumn{6}{c}{Poisson Model}   \\ \midrule 
& \multicolumn{3}{c}{Gamma}  & \multicolumn{3}{c}{Log-Normal} & & \multicolumn{3}{c}{Gamma}  & \multicolumn{3}{c}{Log-Normal}  \\ \midrule 
& $c_v=0.5$ & $c_v=1.0$ & $c_v=2.0$  & $c_v=0.5$ & $c_v=1.0$ & $c_v=2.0$ & & $c_v=0.5$ & $c_v=1.0$ & $c_v=2.0$  &  $c_v=0.5$ & $c_v=1.0$ & $c_v=2.0$  \\     \toprule 
 \\ 
&\multicolumn{13}{c}{\textbf{Uniform Accident Occurrence}}  \\ 
$\mathbf{\rho^\Phi=0.1:}$   &       & &   &     &  & & &       &   &  & & & \\ 
$\xi^{1a}$ &    781.9 &    840.4 &   1007.2 &    787.2 &    838.1 &   1012.7 & &    779.0 &    844.5 &   1019.1 &    787.2 &    850.7 &    997.9 \\ 
$\xi^{1b}$ &    760.3 &    818.7 &    982.2 &    769.6 &    814.9 &    978.3 & &    760.0 &    820.8 &    978.6 &    757.8 &    823.5 &    977.2 \\
\bottomrule 
\end{tabular} 
\begin{tablenotes} 
\small 
\item The Value-at-Risk at level $p=0.95$ of the total loss is approximated using $10,000$ independent samples of $L$. 
\end{tablenotes} 
\end{threeparttable} 
}
\label{tab6}

\end{table}

\begin{table} 
\centering 
\scalebox{0.7}{ 
\begin{threeparttable} 
\caption{Expected Shortfall at Level $p=0.95$ of the Total Loss.} 
\begin{tabular}{rrrrrrrrrrrrrr} \toprule 
& \multicolumn{6}{c}{Binomial Model}  & \quad\quad & \multicolumn{6}{c}{Poisson Model}   \\ \midrule 
& \multicolumn{3}{c}{Gamma}  & \multicolumn{3}{c}{Log-Normal} & & \multicolumn{3}{c}{Gamma}  & \multicolumn{3}{c}{Log-Normal}  \\ \midrule 
& $c_v=0.5$ & $c_v=1.0$ & $c_v=2.0$  & $c_v=0.5$ & $c_v=1.0$ & $c_v=2.0$ & & $c_v=0.5$ & $c_v=1.0$ & $c_v=2.0$  &  $c_v=0.5$ & $c_v=1.0$ & $c_v=2.0$  \\     \toprule 
 \\ 
&\multicolumn{13}{c}{\textbf{Uniform Accident Occurrence}}  \\ 
$\mathbf{\rho^\Phi=0.1:}$   &       & &   &     &  & & &       &   &  & & & \\ 
$\xi^{1a}$ &    851.6 &    934.7 &   1173.4 &    853.3 &    946.7 &   1262.1 & &    845.7 &    931.5 &   1173.7 &    857.9 &    942.1 &   1242.0 \\ 
$\xi^{1b}$ &    823.8 &    904.7 &   1144.5 &    835.0 &    916.0 &   1225.7 & &    825.6 &    908.2 &   1133.5 &    821.0 &    922.4 &   1217.5 \\
\bottomrule 
\end{tabular} 
\begin{tablenotes} 
\small 
\item The Expected Shortfall at level $p=0.95$ of the total loss is approximated using $10,000$ independent samples of $L$. 
\end{tablenotes} 
\end{threeparttable} 
}
\label{tab7}

\end{table}

\begin{table} 
\centering 
\scalebox{0.7}{ 
\begin{threeparttable} 
\caption{Value-at-Risk at Level $p=0.99$ of the Total Loss.} 
\begin{tabular}{rrrrrrrrrrrrrr} \toprule 
& \multicolumn{6}{c}{Binomial Model}  & \quad\quad & \multicolumn{6}{c}{Poisson Model}   \\ \midrule 
& \multicolumn{3}{c}{Gamma}  & \multicolumn{3}{c}{Log-Normal} & & \multicolumn{3}{c}{Gamma}  & \multicolumn{3}{c}{Log-Normal}  \\ \midrule 
& $c_v=0.5$ & $c_v=1.0$ & $c_v=2.0$  & $c_v=0.5$ & $c_v=1.0$ & $c_v=2.0$ & & $c_v=0.5$ & $c_v=1.0$ & $c_v=2.0$  &  $c_v=0.5$ & $c_v=1.0$ & $c_v=2.0$  \\     \toprule 
 \\ 
&\multicolumn{13}{c}{\textbf{Uniform Accident Occurrence}}  \\ 
$\mathbf{\rho^\Phi=0.1:}$   &       & &   &     &  & & &       &   &  & & & \\ 
$\xi^{1a}$ &    896.9 &    993.6 &   1276.6 &    898.9 &   1011.2 &   1411.3 & &    886.1 &    977.0 &   1268.0 &    898.6 &   1000.2 &   1361.8 \\ 
$\xi^{1b}$ &    865.4 &    964.3 &   1243.6 &    877.0 &    978.3 &   1369.8 & &    872.9 &    960.5 &   1221.0 &    861.4 &    989.1 &   1338.5 \\
\bottomrule 
\end{tabular} 
\begin{tablenotes} 
\small 
\item The Value-at-Risk at level $p=0.99$ of the total loss is approximated using $10,000$ independent samples of $L$. 
\end{tablenotes} 
\end{threeparttable} 
}
\label{tab8}

\end{table}

\begin{table} 
\centering 
\scalebox{0.7}{ 
\begin{threeparttable} 
\caption{Expected Shortfall at Level $p=0.99$ of the Total Loss.} 
\begin{tabular}{rrrrrrrrrrrrrr} \toprule 
& \multicolumn{6}{c}{Binomial Model}  & \quad\quad & \multicolumn{6}{c}{Poisson Model}   \\ \midrule 
& \multicolumn{3}{c}{Gamma}  & \multicolumn{3}{c}{Log-Normal} & & \multicolumn{3}{c}{Gamma}  & \multicolumn{3}{c}{Log-Normal}  \\ \midrule 
& $c_v=0.5$ & $c_v=1.0$ & $c_v=2.0$  & $c_v=0.5$ & $c_v=1.0$ & $c_v=2.0$ & & $c_v=0.5$ & $c_v=1.0$ & $c_v=2.0$  &  $c_v=0.5$ & $c_v=1.0$ & $c_v=2.0$  \\     \toprule 
 \\ 
&\multicolumn{13}{c}{\textbf{Uniform Accident Occurrence}}  \\ 
$\mathbf{\rho^\Phi=0.1:}$   &       & &   &     &  & & &       &   &  & & & \\ 
$\xi^{1a}$ &    951.2 &   1084.1 &   1447.0 &    948.7 &   1127.3 &   1695.8 & &    950.2 &   1059.9 &   1431.3 &    961.0 &   1078.2 &   1681.4 \\ 
$\xi^{1b}$ &    915.9 &   1029.1 &   1413.4 &    928.7 &   1073.8 &   1659.5 & &    920.4 &   1030.0 &   1374.7 &    918.9 &   1061.8 &   1648.4 \\
\bottomrule 
\end{tabular} 
\begin{tablenotes} 
\small 
\item The Expected Shortfall at level $p=0.99$ of the total loss is approximated using $10,000$ independent samples of $L$. 
\end{tablenotes} 
\end{threeparttable} 
}
\label{tab9}

\end{table}

\begin{table} 
\centering 
\scalebox{0.7}{ 
\begin{threeparttable} 
\caption{Expectation of the Normalized Total Loss.} 
\begin{tabular}{rrrrrrrrrrrrrr} \toprule 
& \multicolumn{6}{c}{Binomial Model}  & \quad\quad & \multicolumn{6}{c}{Poisson Model}   \\ \midrule 
& \multicolumn{3}{c}{Gamma}  & \multicolumn{3}{c}{Log-Normal} & & \multicolumn{3}{c}{Gamma}  & \multicolumn{3}{c}{Log-Normal}  \\ \midrule 
& $c_v=0.5$ & $c_v=1.0$ & $c_v=2.0$  & $c_v=0.5$ & $c_v=1.0$ & $c_v=2.0$ & & $c_v=0.5$ & $c_v=1.0$ & $c_v=2.0$  &  $c_v=0.5$ & $c_v=1.0$ & $c_v=2.0$  \\     \toprule 
 \\ 
&\multicolumn{13}{c}{\textbf{Uniform Accident Occurrence}}  \\ 
$\mathbf{\rho^\Phi=0.1:}$   &       & &   &     &  & & &       &   &  & & & \\ 
$\xi^{1a}$ &    140.7 &    140.2 &    140.1 &    140.8 &    140.1 &    140.7 & &    140.4 &    140.7 &    140.6 &    139.9 &    140.8 &    139.7 \\ 
$\xi^{1b}$ &    135.7 &    135.4 &    134.6 &    136.2 &    135.1 &    134.4 & &    135.6 &    135.8 &    134.6 &    134.7 &    136.3 &    135.3 \\
\bottomrule 
\end{tabular} 
\begin{tablenotes} 
\small 
\item The expectation of the normalized total loss is approximated using $10,000$ independent samples of $L$. Total losses are normalized by $100$ expected insured vehicles.
\end{tablenotes} 
\end{threeparttable} 
}
\label{tab1}

\end{table}

\begin{table} 
\centering 
\scalebox{0.7}{ 
\begin{threeparttable} 
\caption{Variance of the Normalized Total Loss.} 
\begin{tabular}{rrrrrrrrrrrrrr} \toprule 
& \multicolumn{6}{c}{Binomial Model}  & \quad\quad & \multicolumn{6}{c}{Poisson Model}   \\ \midrule 
& \multicolumn{3}{c}{Gamma}  & \multicolumn{3}{c}{Log-Normal} & & \multicolumn{3}{c}{Gamma}  & \multicolumn{3}{c}{Log-Normal}  \\ \midrule 
& $c_v=0.5$ & $c_v=1.0$ & $c_v=2.0$  & $c_v=0.5$ & $c_v=1.0$ & $c_v=2.0$ & & $c_v=0.5$ & $c_v=1.0$ & $c_v=2.0$  &  $c_v=0.5$ & $c_v=1.0$ & $c_v=2.0$  \\     \toprule 
 \\ 
&\multicolumn{13}{c}{\textbf{Uniform Accident Occurrence}}  \\ 
$\mathbf{\rho^\Phi=0.1:}$   &       & &   &     &  & & &       &   &  & & & \\ 
$\xi^{1a}$ &   1251.8 &   1833.0 &   4195.9 &   1245.5 &   1856.9 &   4339.0 & &   1211.0 &   1860.9 &   4190.3 &   1263.6 &   1861.2 &   4170.3 \\ 
$\xi^{1b}$ &   1198.4 &   1772.0 &   4018.9 &   1221.8 &   1729.8 &   4092.4 & &   1185.1 &   1776.4 &   4028.5 &   1165.4 &   1809.5 &   4043.0 \\
\bottomrule 
\end{tabular} 
\begin{tablenotes} 
\small 
\item The variance of the normalized total loss is approximated using $10,000$ independent samples of $L$. Total losses are normalized by $100$ expected insured vehicles.
\end{tablenotes} 
\end{threeparttable} 
}
\label{tab2}

\end{table}

\begin{table} 
\centering 
\scalebox{0.7}{ 
\begin{threeparttable} 
\caption{Skewness of the Normalized Total Loss.} 
\begin{tabular}{rrrrrrrrrrrrrr} \toprule 
& \multicolumn{6}{c}{Binomial Model}  & \quad\quad & \multicolumn{6}{c}{Poisson Model}   \\ \midrule 
& \multicolumn{3}{c}{Gamma}  & \multicolumn{3}{c}{Log-Normal} & & \multicolumn{3}{c}{Gamma}  & \multicolumn{3}{c}{Log-Normal}  \\ \midrule 
& $c_v=0.5$ & $c_v=1.0$ & $c_v=2.0$  & $c_v=0.5$ & $c_v=1.0$ & $c_v=2.0$ & & $c_v=0.5$ & $c_v=1.0$ & $c_v=2.0$  &  $c_v=0.5$ & $c_v=1.0$ & $c_v=2.0$  \\     \toprule 
 \\ 
&\multicolumn{13}{c}{\textbf{Uniform Accident Occurrence}}  \\ 
$\mathbf{\rho^\Phi=0.1:}$   &       & &   &     &  & & &       &   &  & & & \\ 
$\xi^{1a}$ &    0.306 &    0.533 &    0.869 &    0.320 &    0.679 &    1.904 & &    0.327 &    0.454 &    0.858 &    0.386 &    0.536 &    1.728 \\ 
$\xi^{1b}$ &    0.292 &    0.468 &    0.868 &    0.357 &    0.666 &    1.764 & &    0.330 &    0.468 &    0.818 &    0.366 &    0.563 &    1.756 \\
\bottomrule 
\end{tabular} 
\begin{tablenotes} 
\small 
\item The skewness of the normalized total loss is approximated using $10,000$ independent samples of $L$. Total losses are normalized by $100$ expected insured vehicles.
\end{tablenotes} 
\end{threeparttable} 
}
\label{tab3}

\end{table}

\begin{table} 
\centering 
\scalebox{0.7}{ 
\begin{threeparttable} 
\caption{Value-at-Risk at Level $p=0.9$ of the Normalized Total Loss.} 
\begin{tabular}{rrrrrrrrrrrrrr} \toprule 
& \multicolumn{6}{c}{Binomial Model}  & \quad\quad & \multicolumn{6}{c}{Poisson Model}   \\ \midrule 
& \multicolumn{3}{c}{Gamma}  & \multicolumn{3}{c}{Log-Normal} & & \multicolumn{3}{c}{Gamma}  & \multicolumn{3}{c}{Log-Normal}  \\ \midrule 
& $c_v=0.5$ & $c_v=1.0$ & $c_v=2.0$  & $c_v=0.5$ & $c_v=1.0$ & $c_v=2.0$ & & $c_v=0.5$ & $c_v=1.0$ & $c_v=2.0$  &  $c_v=0.5$ & $c_v=1.0$ & $c_v=2.0$  \\     \toprule 
 \\ 
&\multicolumn{13}{c}{\textbf{Uniform Accident Occurrence}}  \\ 
$\mathbf{\rho^\Phi=0.1:}$   &       & &   &     &  & & &       &   &  & & & \\ 
$\xi^{1a}$ &    187.1 &    197.2 &    227.2 &    187.3 &    196.4 &    221.7 & &    186.7 &    198.9 &    228.3 &    186.1 &    198.7 &    219.2 \\ 
$\xi^{1b}$ &    182.2 &    191.4 &    218.3 &    182.6 &    189.9 &    213.5 & &    181.7 &    191.3 &    220.7 &    179.9 &    192.7 &    213.0 \\
\bottomrule 
\end{tabular} 
\begin{tablenotes} 
\small 
\item The Value-at-Risk at level $p=0.9$ of the normalized total loss is approximated using $10,000$ independent samples of $L$. Total losses are normalized by $100$ expected insured vehicles.
\end{tablenotes} 
\end{threeparttable} 
}
\label{tab4}

\end{table}

\begin{table} 
\centering 
\scalebox{0.7}{ 
\begin{threeparttable} 
\caption{Expected Shortfall at Level $p=0.9$ of the Normalized Total Loss.} 
\begin{tabular}{rrrrrrrrrrrrrr} \toprule 
& \multicolumn{6}{c}{Binomial Model}  & \quad\quad & \multicolumn{6}{c}{Poisson Model}   \\ \midrule 
& \multicolumn{3}{c}{Gamma}  & \multicolumn{3}{c}{Log-Normal} & & \multicolumn{3}{c}{Gamma}  & \multicolumn{3}{c}{Log-Normal}  \\ \midrule 
& $c_v=0.5$ & $c_v=1.0$ & $c_v=2.0$  & $c_v=0.5$ & $c_v=1.0$ & $c_v=2.0$ & & $c_v=0.5$ & $c_v=1.0$ & $c_v=2.0$  &  $c_v=0.5$ & $c_v=1.0$ & $c_v=2.0$  \\     \toprule 
 \\ 
&\multicolumn{13}{c}{\textbf{Uniform Accident Occurrence}}  \\ 
$\mathbf{\rho^\Phi=0.1:}$   &       & &   &     &  & & &       &   &  & & & \\ 
$\xi^{1a}$ &    206.4 &    223.0 &    271.7 &    206.9 &    224.4 &    282.2 & &    205.3 &    223.4 &    272.5 &    207.3 &    225.4 &    277.7 \\ 
$\xi^{1b}$ &    200.3 &    216.3 &    263.9 &    202.5 &    217.4 &    272.4 & &    200.1 &    216.9 &    263.3 &    199.1 &    219.7 &    271.4 \\
\bottomrule 
\end{tabular} 
\begin{tablenotes} 
\small 
\item The Expected Shortfall at level $p=0.9$ of the normalized total loss is approximated using $10,000$ independent samples of $L$. Total losses are normalized by $100$ expected insured vehicles.
\end{tablenotes} 
\end{threeparttable} 
}
\label{tab5}

\end{table}

\begin{table} 
\centering 
\scalebox{0.7}{ 
\begin{threeparttable} 
\caption{Value-at-Risk at Level $p=0.95$ of the Normalized Total Loss.} 
\begin{tabular}{rrrrrrrrrrrrrr} \toprule 
& \multicolumn{6}{c}{Binomial Model}  & \quad\quad & \multicolumn{6}{c}{Poisson Model}   \\ \midrule 
& \multicolumn{3}{c}{Gamma}  & \multicolumn{3}{c}{Log-Normal} & & \multicolumn{3}{c}{Gamma}  & \multicolumn{3}{c}{Log-Normal}  \\ \midrule 
& $c_v=0.5$ & $c_v=1.0$ & $c_v=2.0$  & $c_v=0.5$ & $c_v=1.0$ & $c_v=2.0$ & & $c_v=0.5$ & $c_v=1.0$ & $c_v=2.0$  &  $c_v=0.5$ & $c_v=1.0$ & $c_v=2.0$  \\     \toprule 
 \\ 
&\multicolumn{13}{c}{\textbf{Uniform Accident Occurrence}}  \\ 
$\mathbf{\rho^\Phi=0.1:}$   &       & &   &     &  & & &       &   &  & & & \\ 
$\xi^{1a}$ &    201.2 &    216.3 &    259.3 &    202.6 &    215.7 &    260.7 & &    200.5 &    217.4 &    262.3 &    202.6 &    219.0 &    256.9 \\ 
$\xi^{1b}$ &    195.7 &    210.7 &    252.8 &    198.1 &    209.7 &    251.8 & &    195.6 &    211.3 &    251.9 &    195.1 &    212.0 &    251.5 \\
\bottomrule 
\end{tabular} 
\begin{tablenotes} 
\small 
\item The Value-at-Risk at level $p=0.95$ of the normalized total loss is approximated using $10,000$ independent samples of $L$. Total losses are normalized by $100$ expected insured vehicles.
\end{tablenotes} 
\end{threeparttable} 
}
\label{tab6}

\end{table}

\begin{table} 
\centering 
\scalebox{0.7}{ 
\begin{threeparttable} 
\caption{Expected Shortfall at Level $p=0.95$ of the Normalized Total Loss.} 
\begin{tabular}{rrrrrrrrrrrrrr} \toprule 
& \multicolumn{6}{c}{Binomial Model}  & \quad\quad & \multicolumn{6}{c}{Poisson Model}   \\ \midrule 
& \multicolumn{3}{c}{Gamma}  & \multicolumn{3}{c}{Log-Normal} & & \multicolumn{3}{c}{Gamma}  & \multicolumn{3}{c}{Log-Normal}  \\ \midrule 
& $c_v=0.5$ & $c_v=1.0$ & $c_v=2.0$  & $c_v=0.5$ & $c_v=1.0$ & $c_v=2.0$ & & $c_v=0.5$ & $c_v=1.0$ & $c_v=2.0$  &  $c_v=0.5$ & $c_v=1.0$ & $c_v=2.0$  \\     \toprule 
 \\ 
&\multicolumn{13}{c}{\textbf{Uniform Accident Occurrence}}  \\ 
$\mathbf{\rho^\Phi=0.1:}$   &       & &   &     &  & & &       &   &  & & & \\ 
$\xi^{1a}$ &    219.2 &    240.6 &    302.0 &    219.6 &    243.7 &    324.9 & &    217.7 &    239.8 &    302.1 &    220.8 &    242.5 &    319.7 \\ 
$\xi^{1b}$ &    212.0 &    232.9 &    294.6 &    214.9 &    235.8 &    315.5 & &    212.5 &    233.8 &    291.8 &    211.3 &    237.4 &    313.4 \\
\bottomrule 
\end{tabular} 
\begin{tablenotes} 
\small 
\item The Expected Shortfall at level $p=0.95$ of the normalized total loss is approximated using $10,000$ independent samples of $L$. Total losses are normalized by $100$ expected insured vehicles.
\end{tablenotes} 
\end{threeparttable} 
}
\label{tab7}

\end{table}

\begin{table} 
\centering 
\scalebox{0.7}{ 
\begin{threeparttable} 
\caption{Value-at-Risk at Level $p=0.99$ of the Normalied Total Loss.} 
\begin{tabular}{rrrrrrrrrrrrrr} \toprule 
& \multicolumn{6}{c}{Binomial Model}  & \quad\quad & \multicolumn{6}{c}{Poisson Model}   \\ \midrule 
& \multicolumn{3}{c}{Gamma}  & \multicolumn{3}{c}{Log-Normal} & & \multicolumn{3}{c}{Gamma}  & \multicolumn{3}{c}{Log-Normal}  \\ \midrule 
& $c_v=0.5$ & $c_v=1.0$ & $c_v=2.0$  & $c_v=0.5$ & $c_v=1.0$ & $c_v=2.0$ & & $c_v=0.5$ & $c_v=1.0$ & $c_v=2.0$  &  $c_v=0.5$ & $c_v=1.0$ & $c_v=2.0$  \\     \toprule 
 \\ 
&\multicolumn{13}{c}{\textbf{Uniform Accident Occurrence}}  \\ 
$\mathbf{\rho^\Phi=0.1:}$   &       & &   &     &  & & &       &   &  & & & \\ 
$\xi^{1a}$ &    230.9 &    255.8 &    328.6 &    231.4 &    260.3 &    363.3 & &    228.1 &    251.5 &    326.4 &    231.3 &    257.4 &    350.5 \\ 
$\xi^{1b}$ &    222.8 &    248.2 &    320.1 &    225.7 &    251.8 &    352.6 & &    224.7 &    247.2 &    314.3 &    221.7 &    254.6 &    344.5 \\
\bottomrule 
\end{tabular} 
\begin{tablenotes} 
\small 
\item The Value-at-Risk at level $p=0.99$ of the normalized total loss is approximated using $10,000$ independent samples of $L$. Total losses are normalized by $100$ expected insured vehicles.
\end{tablenotes} 
\end{threeparttable} 
}
\label{tab8}

\end{table}

\begin{table} 
\centering 
\scalebox{0.7}{ 
\begin{threeparttable} 
\caption{Expected Shortfall at Level $p=0.99$ of the Normalized Total Loss.} 
\begin{tabular}{rrrrrrrrrrrrrr} \toprule 
& \multicolumn{6}{c}{Binomial Model}  & \quad\quad & \multicolumn{6}{c}{Poisson Model}   \\ \midrule 
& \multicolumn{3}{c}{Gamma}  & \multicolumn{3}{c}{Log-Normal} & & \multicolumn{3}{c}{Gamma}  & \multicolumn{3}{c}{Log-Normal}  \\ \midrule 
& $c_v=0.5$ & $c_v=1.0$ & $c_v=2.0$  & $c_v=0.5$ & $c_v=1.0$ & $c_v=2.0$ & & $c_v=0.5$ & $c_v=1.0$ & $c_v=2.0$  &  $c_v=0.5$ & $c_v=1.0$ & $c_v=2.0$  \\     \toprule 
 \\ 
&\multicolumn{13}{c}{\textbf{Uniform Accident Occurrence}}  \\ 
$\mathbf{\rho^\Phi=0.1:}$   &       & &   &     &  & & &       &   &  & & & \\ 
$\xi^{1a}$ &    244.8 &    279.0 &    372.4 &    244.2 &    290.2 &    436.5 & &    244.6 &    272.8 &    368.4 &    247.4 &    277.5 &    432.8 \\ 
$\xi^{1b}$ &    235.8 &    264.9 &    363.8 &    239.0 &    276.4 &    427.1 & &    236.9 &    265.1 &    353.9 &    236.5 &    273.3 &    424.3 \\
\bottomrule 
\end{tabular} 
\begin{tablenotes} 
\small 
\item The Expected Shortfall at level $p=0.99$ of the normalized total loss is approximated using $10,000$ independent samples of $L$. Total losses are normalized by $100$ expected insured vehicles.
\end{tablenotes} 
\end{threeparttable} 
}
\label{tab9}

\end{table}

\end{document}